\documentclass[journal]{IEEEtran}
\usepackage{dblfloatfix}
\usepackage{placeins}
\usepackage{amsmath,amsfonts}
\usepackage{graphicx}
\usepackage{float}
\usepackage[hidelinks]{hyperref}
\usepackage{cite}
\usepackage{braket}
\usepackage{url}
\graphicspath{{./}}

\begin{document}

\title{Parameterized Quantum Circuits as Feature Maps: Representation Quality and Readout Effects in Multispectral Land-Cover Classification}

\author{
Ralntion Komini,
Aikaterini Mandilara,
Georgios Maragkopoulos,
and Dimitris Syvridis
\thanks{R. Komini, A. Mandilara and G. Maragkopoulos are with the Department of Informatics and Telecommunications, National and Kapodistrian University of Athens, Greece (e-mail: raldikomini02@gmail.com; mandkat@di.uoa.gr; giorgosmarag@di.uoa.gr). Corresponding author: R. Komini (e-mail: raldikomini02@gmail.com).}
\thanks{D. Syvridis is with the Department of Informatics and Telecommunications, National and Kapodistrian University of Athens, Greece (e-mail: dsyvridi@di.uoa.gr).}
\thanks{All authors are also with Eulambia Advanced Technologies Ltd, Agiou Ioannou 24, Ag. Paraskevi, 15342, Greece.}
}

\maketitle

\begin{abstract}
\bfseries

Parameterized quantum circuits (PQCs) are increasingly explored for machine learning, yet it remains unclear whether their predictive performance is determined primarily by the learned quantum representation or by the readout used to extract information from it. We address this question for multispectral land-cover classification by treating variational quantum classifiers (VQCs) as learned feature maps rather than standalone classifiers. Using EuroSAT-MS, we perform a controlled evaluation over all 45 one-versus-one class pairs with fixed train-validation-test splits, validation-based model selection, and five optimization seeds. Reusing the trained circuit as a fidelity kernel consistently improves over the standard local VQC readout, indicating that the learned quantum representation contains discriminative information not fully exploited by local measurements. Classical RBF support-vector machines nevertheless remain stronger overall, while image-based ResNet-18 models substantially outperform all compact PCA-based approaches by exploiting spatial information discarded during preprocessing. Additional experiments on So2Sat-LCZ42 and SAT-6 show that the relative benefit of different readout strategies is dataset dependent. Rather than demonstrating quantum advantage, our results show that compact PQCs can learn meaningful nonlinear representations of multispectral data and suggest that representation learning, rather than classifier accuracy alone, provides a useful perspective for evaluating near-term quantum machine learning models.

\end{abstract}


\section{Introduction}

Land-cover classification from multispectral satellite imagery is a central problem in Earth observation, with applications in environmental monitoring, urban planning, land-use analysis, and agricultural mapping. Classical machine learning methods such as logistic regression, random forests, and support-vector machines (SVMs) have long been used for this task, while convolutional neural networks (CNNs), transformers, and self-supervised geospatial models now provide strong image-based baselines for large remote-sensing datasets \cite{ma2019dl_remote_sensing_review,helber2019eurosat,sumbul2019bigearthnet,cong2022satmae}.

Quantum machine learning (QML) is being explored in a very different computational regime. Present-day quantum processors provide only a limited number of reliable qubits and can execute only shallow circuits before noise becomes dominant. Under these constraints, it is premature to frame small variational quantum classifiers as replacements for mature classical vision architectures. A more informative question is whether a compact parameterized quantum circuit (PQC) can learn a useful nonlinear representation of multispectral inputs, and how much of that representation is accessible to different readout mechanisms.

This paper adopts that feature-map perspective. A VQC maps a classical input \(\boldsymbol{x}\) to a quantum state \(\ket{\psi_{\boldsymbol{\theta}}(\boldsymbol{x})}\), after which a measurement and classical decision rule are used for classification. The circuit therefore plays the role of a learned nonlinear embedding, while the readout determines how the embedded representation is used. This interpretation is consistent with the view of quantum models as feature maps in Hilbert space \cite{schuld2019feature_hilbert_spaces,havlivcek2019quantum_enhanced_feature_spaces,schuld2021qml_kernel_methods,schuld2021advantages} and with recent analyses of PQC expressivity \cite{cerezo2021vqa_review,abbas2021universal}.

The distinction between representation and readout is important. A standard VQC with a linear head observes only a small set of local expectation values, whereas a quantum-kernel SVM can use the pairwise state-overlap geometry induced by the same trained circuit. The two models therefore share the same learned quantum feature map but access it differently. This separation allows us to ask whether the trained circuit contains discriminative structure that is not fully exploited by the usual local linear measurement head.

We study this question primarily on EuroSAT-MS \cite{helber2018introducing,helber2019eurosat}. The main evaluation uses all 45 one-vs-one class pairs, with fixed train-validation-test splits, five independent training seeds, and preprocessing fitted only on the training partition. Model-selection choices, including PCA dimensionality and circuit architecture, are made using validation data only. The selected configuration is then evaluated on held-out test data and extended to full 10-class one-vs-one voting. To examine whether the same conclusions persist outside EuroSAT, the selected configuration is also run without re-selection on So2Sat-LCZ42 \cite{zhu2020so2sat} and SAT-6 \cite{basu2015deepsat}.

Specifically, this work makes three contributions. First, it provides a controlled pairwise and multiclass evaluation of compact PQC feature maps for multispectral land-cover classification. Second, it separates the effect of the trained quantum representation from the effect of the readout, comparing a local linear VQC head with a trained quantum-kernel SVM constructed from the same circuit. Third, it analyzes the selected representation beyond aggregate accuracy, studying its sensitivity to PCA dimension and circuit architecture, its behavior under increasing qubit count, its computational cost, and its robustness across additional remote-sensing datasets.

The results should not be interpreted as evidence of quantum advantage. Although the strongest classical nonlinear baseline (SVM-RBF) remains superior on the evaluated tasks, the trained quantum kernel consistently improves upon the standard local VQC readout, indicating that the learned PQC representation contains discriminative information that is not fully accessible through local measurements alone. These findings suggest that the principal value of compact PQCs in this setting lies in learning meaningful nonlinear representations, while highlighting the central role of the downstream readout—and its computational cost—in determining their practical usefulness.

\FloatBarrier

\section{Related Work}

Land-cover classification from satellite imagery has been widely studied using both classical machine learning and modern deep learning methods. Traditional approaches often rely on spectral or reduced-dimensional features combined with classifiers such as logistic regression, random forests, and support vector machines. In particular, SVMs remain strong baselines in remote sensing because they can perform well with limited labeled data and can model nonlinear decision boundaries through kernels such as the radial basis function (RBF) kernel. With the growth of labeled remote-sensing benchmarks, deep convolutional neural networks (CNNs) became dominant for land-use and land-cover classification, including on EuroSAT, where deep CNN baselines were evaluated on Sentinel-2 imagery \cite{helber2019eurosat}. Larger-scale datasets such as BigEarthNet further enabled training and benchmarking deep models on multispectral Sentinel-2 image archives \cite{sumbul2019bigearthnet}. More recently, transformer-based and self-supervised approaches have become increasingly important in remote sensing. Vision Transformer and Swin Transformer models have been applied to LULC classification with transfer learning and fine-tuning \cite{khan2024transformer_lulc}, while masked-autoencoder pretraining methods such as SatMAE explicitly exploit temporal and multispectral structure in satellite imagery \cite{cong2022satmae}. Geospatial foundation models such as Prithvi-EO further represent the current trend toward large pretrained models that can be adapted to multiple Earth-observation tasks \cite{szwarcman2025prithvi}. In contrast to these large classical models, our study focuses on compact PQC-based feature maps and compares them against lightweight but carefully controlled classical baselines under identical preprocessing and data splits. This controlled setting allows differences in performance to be attributed to the learned representations rather than to differences in data preprocessing or model selection.

Within QML for Earth observation and remote sensing, prior work has explored kernel-based, hybrid quantum--classical, and quantum-inspired approaches. Quantum SVM methods have been investigated for remote-sensing data classification, including quantum-kernel and quantum-annealing formulations \cite{Delilbasic2021Quantum,Delilbasic2023A,Zollner2022Quantum,Miroszewski2023Cloud}. Circuit-based hybrid quantum NNs and quanvolutional models have also been applied to Earth-observation imagery, including multispectral and satellite-image classification tasks \cite{Sebastianelli2021On,Fan2023Hybrid,Fan2024Land,Liliopoulos2025Hybrid,Sebastianelli2024Quanv4EO}. Related work has further examined quantum hyperparameter selection, practical bottlenecks, and the broader potential of QML for remote sensing \cite{Zaidenberg2021Advantages,Sebastianelli2023On,Otgonbaatar2023Exploiting,Liu2023Quantum,Shaik2022Quantum}. In parallel, quantum-inspired architectures based on unitary or Hilbert-space transformations have been proposed for multispectral satellite classification \cite{maragkopoulos2026quantum}.  A closely related quantum-inspired Hilbert-space approach is also considered in \cite{katerina}.  Unlike previous studies, which primarily evaluate end-to-end classification performance, our work interprets the trained PQC as a learned nonlinear feature map and investigates how different readout mechanisms—a local measurement head and a trained quantum-kernel SVM—access the same learned representation.

Accordingly, the present work differs from previous studies in two important respects. First, it does not evaluate only the final VQC accuracy. Instead, it treats the trained PQC as a feature map and compares how different readouts exploit the same learned quantum representation. Second, it separates compact PCA-based spectral experiments from direct image-based CNN baselines. The PCA-based models provide a controlled setting for comparing VQC, quantum-kernel, and classical-kernel behavior under identical low-dimensional inputs, while the ResNet baselines show how these compact representations compare with standard image models that use spatial structure directly.

\FloatBarrier

\section{Parameterized Quantum Circuits as Feature Maps}
\label{sec:theory}

A VQC consists of a data-encoding circuit, trainable quantum gates, an entangling pattern, a measurement rule, and a classical loss optimized by a classical algorithm \cite{Benedetti2019PQC}. Given a preprocessed input vector \(\boldsymbol{x}\), the circuit prepares a state
\begin{equation}
\ket{\psi_{\boldsymbol{\theta}}(\boldsymbol{x})}
=
U_{\boldsymbol{\theta}}(\boldsymbol{x})\ket{0}^{\otimes q},
\label{eq:state}
\end{equation}
where \(q\) is the number of qubits and \(\boldsymbol{\theta}\) denotes trainable parameters. In data re-uploading circuits, \(\boldsymbol{x}\) enters the circuit repeatedly across multiple layers, making the state-preparation unitary both input-dependent and trainable. Consequently, the trainable parameters determine not only the classifier but also the geometry of the embedding of classical data into Hilbert space.

\begin{figure*}[!t]

\centering
\includegraphics[width=\textwidth]{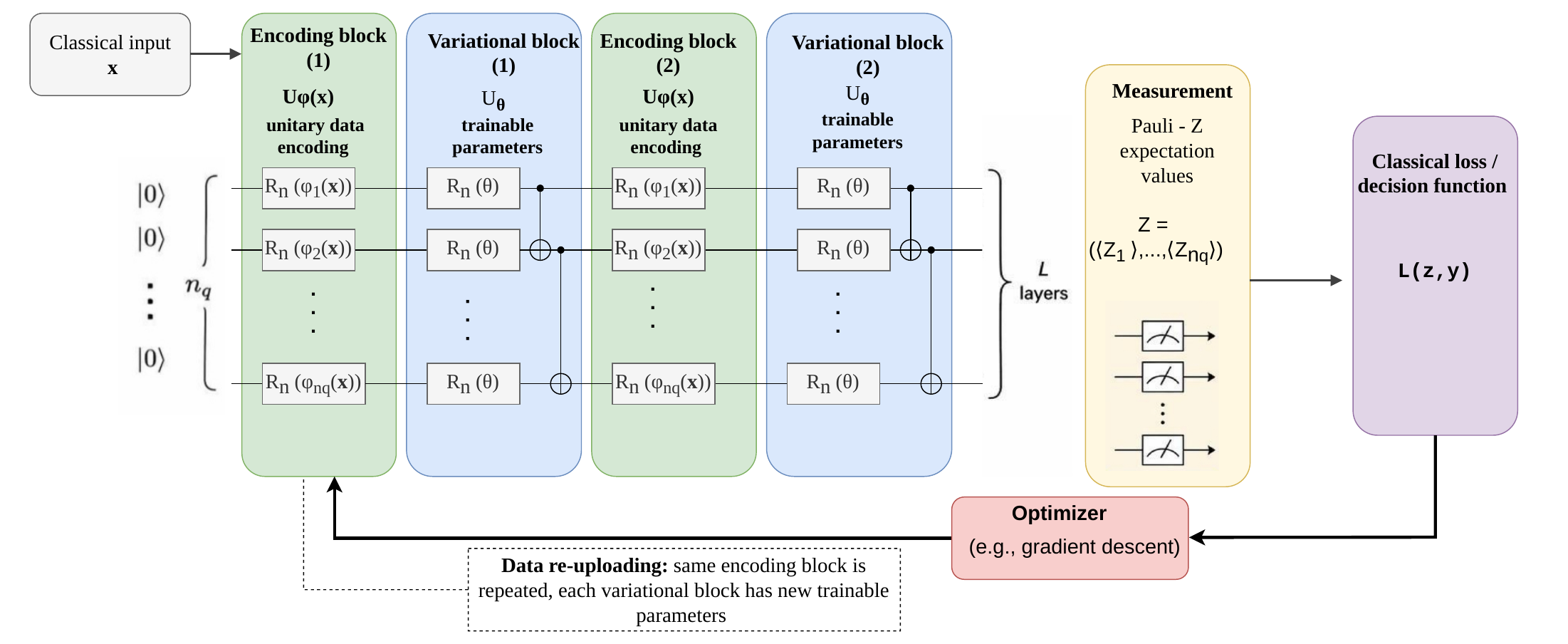}
\caption{Generic VQC architecture with data reuploading. A classical input \(\boldsymbol{x}\) is encoded through data-dependent rotations, followed by trainable single-qubit gates and entangling gates. The process is repeated for \(L\) blocks. Local Pauli-\(Z\) expectation values are measured and passed to a classical decision rule. The same trained circuit can also be frozen and reused as a quantum feature map for a fidelity-kernel SVM.}
\label{fig:vqc_generic}
\end{figure*}

Figure~\ref{fig:vqc_generic} illustrates the generic VQC architecture considered in this work and summarizes the two complementary viewpoints developed throughout this section. The same trained PQC is evaluated both with the standard linear measurement head and with a quantum-kernel SVM, allowing the effect of the readout to be separated from that of the learned quantum representation. The following subsections formalize these two perspectives.

\subsection{\texorpdfstring{VQC with a Linear Measurement Head}{VQC With a Linear Measurement Head}}

Although the quantum state of Eq.~(\ref{eq:state}) resides in a high-dimensional Hilbert space, the standard VQC prediction is constructed from only q measured expectation values. Consequently, the classifier has direct access only to this low-dimensional vector of observables. Any additional information encoded in the quantum state can influence the prediction only insofar as it is mapped onto these measured quantities during training.

In the standard VQC readout, also adopted in this work, one measures local Pauli-\(Z\) expectation values. Let
\begin{equation}
Z_j = I_1 \otimes \cdots \otimes I_{j-1} \otimes Z_j \otimes I_{j+1} \otimes \cdots \otimes I_q
\end{equation}
denote the Pauli-\(Z\) operator acting on qubit \(j\). 

Using the corresponding density operator
\(\rho_{\boldsymbol{\theta}}(\boldsymbol{x})
=
\ket{\psi_{\boldsymbol{\theta}}(\boldsymbol{x})}
\bra{\psi_{\boldsymbol{\theta}}(\boldsymbol{x})}\),
the measured feature vector is 
\begin{equation}
\boldsymbol{z}_{\boldsymbol{\theta}}(\boldsymbol{x})
=
\left(
\mathrm{Tr}[\rho_{\boldsymbol{\theta}}(\boldsymbol{x})Z_1],
\ldots,
\mathrm{Tr}[\rho_{\boldsymbol{\theta}}(\boldsymbol{x})Z_q]
\right).
\end{equation}
This feature vector is subsequently passed to a classical linear head, yielding the binary decision function
\begin{equation}
f_{\mathrm{lin}}(\boldsymbol{x})
=
\boldsymbol{w}^{T}\boldsymbol{z}_{\boldsymbol{\theta}}(\boldsymbol{x})+b
=
\mathrm{Tr}\!\left[
\rho_{\boldsymbol{\theta}}(\boldsymbol{x})(O_{\boldsymbol{w}}+bI)
\right],
\label{eq:vqc-linear}
\end{equation}
where \(O_{\boldsymbol{w}}=\sum_{j=1}^{q}w_j Z_j\), \(I\) is the \(2^q\times2^q\) identity, and the bias term is written inside the trace using \(\mathrm{Tr}[\rho_{\boldsymbol{\theta}}(\boldsymbol{x})]=1\).

This readout is simple and naturally compatible with near-term quantum hardware, but it directly exploits only a $q$-dimensional vector of local expectation values. Information encoded elsewhere in the quantum state can contribute to the decision only if training transfers it onto these measured observables. This raises the question of whether alternative readout strategies can extract additional information from the same learned quantum representation.

\subsection{\texorpdfstring{Quantum-Kernel Readout of the Learned Feature Map}{Quantum-Kernel Readout}}

The learned quantum representation can be exploited in ways other than the standard linear measurement head. In particular, once the VQC has been trained, the resulting state preparation circuit defines a parameterized quantum feature map whose output states can be compared via their pairwise fidelities. This naturally induces the quantum kernel
\begin{equation}
K_{\boldsymbol{\theta}}(\boldsymbol{x},\boldsymbol{x}')
=
\left|
\left\langle
\psi_{\boldsymbol{\theta}}(\boldsymbol{x})
\middle|
\psi_{\boldsymbol{\theta}}(\boldsymbol{x}')
\right\rangle
\right|^2
=
\mathrm{Tr}\!\left[
\rho_{\boldsymbol{\theta}}(\boldsymbol{x})
\rho_{\boldsymbol{\theta}}(\boldsymbol{x}')
\right].
\label{eq:qk}
\end{equation}
Unlike conventional quantum-kernel methods, which employ the quantum feature map directly to construct the kernel, we first train the PQC using the standard linear VQC head. The learned circuit is then frozen, the linear head is discarded, and the resulting quantum states are used to construct the fidelity kernel of Eq.~(\ref{eq:qk}).

More specifically, after VQC training, \(\boldsymbol{\theta}\) is frozen and an SVM is trained on the Gram matrix induced by \(K_{\boldsymbol{\theta}}\). The resulting decision function can be written as
\begin{equation}
f_{\mathrm{QK}}(\boldsymbol{x})
=
\sum_{i=1}^{N}\alpha_i y_i
K_{\boldsymbol{\theta}}(\boldsymbol{x}_i,\boldsymbol{x})
+\beta,
\label{eq:qk-svm}
\end{equation}
where \(\boldsymbol{x}_i\) are training samples and \(\alpha_i\) are the learned SVM dual coefficients. Equivalently,
\begin{equation}
f_{\mathrm{QK}}(\boldsymbol{x})
=
\mathrm{Tr}\!\left[
\rho_{\boldsymbol{\theta}}(\boldsymbol{x})
O_{\mathrm{SVM}}
\right]+\beta,
\end{equation}
where we have defined the observable
\begin{equation}
O_{\mathrm{SVM}}
=
\sum_{i=1}^{N}\alpha_i y_i
\rho_{\boldsymbol{\theta}}(\boldsymbol{x}_i).
\label{eq:qk-observable}
\end{equation}

Equation~(\ref{eq:qk-observable}) provides an alternative physical interpretation of the kernel classifier. Rather than viewing the SVM simply as a classical post-processing step, its prediction can again be expressed as the expectation value of an observable. In contrast to the linear VQC head, however, this observable is constructed from the training states themselves and therefore adapts to the learned geometry of the quantum feature map.

Consequently, the kernel readout is not restricted to a fixed linear combination of local Pauli-Z operators, but instead probes the learned representation through a learned, data-dependent observable.

\subsection{\texorpdfstring{Training-Induced Changes in the Kernel}{Why Training Can Change the Kernel}}

At first sight, it may appear that training cannot change pairwise state fidelities because unitary transformations preserve inner products. This intuition would be correct if every input were acted upon by the same unitary operator. data re-uploading circuits are fundamentally different: the implemented unitary \(U_{\boldsymbol{\theta}}(\boldsymbol{x})\) depends explicitly on the input. Training therefore modifies the input-dependent embedding, allowing the geometry of the learned feature map—and consequently the induced kernel—to evolve during optimization.

The linear VQC head accesses this representation only through a fixed set of local expectation values, whereas the kernel SVM bases its decision on pairwise similarities between the embedded training states. Consequently, if training increases within-class fidelities while reducing between-class fidelities, the kernel classifier can exploit this learned geometry more effectively than a linear decision function acting on local observables. Whether such a geometry actually emerges, however, is an empirical question rather than a theoretical guarantee.

The two readout strategies therefore interrogate the same learned quantum representation in fundamentally different ways. The central question addressed in this paper is not whether training changes the representation—it does—but whether the richer kernel-based readout can exploit that representation more effectively under a controlled remote-sensing protocol.

\FloatBarrier

\section{\texorpdfstring{Experimental Protocol}{Experimental Protocol}}

\label{sec:protocol}

\subsection{\texorpdfstring{Datasets and Data Splits}{Dataset and Splits}}

Experiments use the multispectral version of EuroSAT, which contains Sentinel-2 image patches from 10 land-cover classes \cite{helber2018introducing,helber2019eurosat}: AnnualCrop, Forest, HerbaceousVegetation, Highway, Industrial, Pasture, PermanentCrop, Residential, River, and SeaLake. The pairwise experiments consider all \(10 \choose 2\) \(=45\) one-vs-one class pairs. For each class, a fixed global split is formed with 1400 training samples, 300 validation samples, and 300 test samples. Thus, the 10-class split contains 14000 training, 3000 validation, and 3000 test images. The same image identifiers are used for all models. Training seeds affect only optimization, initialization, and mini-batch order; they do not change the data partition.

For a binary pair, the training set therefore contains 2800 samples, while validation and test contain 600 samples each. For the multiclass one-vs-one evaluation, the same trained pairwise classifiers score the common 3000-sample global test partition, and votes are aggregated over all 45 classifiers. Ties are resolved using classifier margins followed by a fixed class order. The validation set is used exclusively for model selection, while the test set is reserved for the final evaluation.

Two additional datasets are used only after the EuroSAT model-selection step. They play no role in selecting the PCA dimension, circuit depth, entangling topology, offset schedule, or any other model hyperparameter. So2Sat-LCZ42 contains Sentinel-1 and Sentinel-2 patches labeled by local climate zone (LCZ) classes across global urban agglomerations \cite{zhu2020so2sat}. To keep the input modality comparable to EuroSAT-MS, we use the Sentinel-2 optical channels only. The experiment uses the official training, validation, and testing files, with balanced subsets of 1400 training, 200 validation, and 200 test samples per class for the 17 LCZ classes. This gives \(17 \choose 2\) \(=136\) one-vs-one pairs and a common 3400-sample global test set for multiclass voting.

SAT-6 is a six-class airborne land-cover dataset with \(28\times28\) RGB+NIR image patches \cite{basu2015deepsat}. It is closer to EuroSAT in label semantics because the classes are land-cover categories rather than LCZ morphology classes, but it is visually simpler and contains fewer classes. We use the official train/test split. For each class, 1400 samples from the official training split are used for training, 300 disjoint samples from the same official training split are used for validation, and 300 samples from the official test split are used for final testing. This gives 15 one-vs-one pairs and a common 1800-sample global test set for multiclass voting.

The external datasets follow the same leakage controls as EuroSAT: PCA and normalization are fitted only on each pair's training subset, then applied unchanged to validation, binary test, and global-test samples. For both external checks, the fixed final EuroSAT configuration is used: PCA-32, \(q=4\), \(L=10\), offset 8, and alternating entanglement. The purpose is therefore to validate the protocol selected on EuroSAT rather than to perform a new dataset-specific hyperparameter search.

This split design is important for two reasons. First, every binary classifier for a given class pair is trained and evaluated on exactly the same image identifiers across VQC, SVM with quantum kernel (SVM-QK), logistic regression, SVM-linear, SVM-RBF, and NN baselines. Pairwise differences therefore reflect model behavior rather than changes in sampled data. Second, the multiclass one-vs-one evaluation is built from pairwise models that all score the same global test images. Consequently, every multiclass prediction is obtained by aggregating the votes of the corresponding 45 pairwise classifiers evaluated on the same global test partition, avoiding inconsistencies that could arise from pair-specific test splits.

\subsection{\texorpdfstring{PCA-Based Leakage-Safe Preprocessing}{Leakage-Safe Preprocessing}}

For the PCA-based experiments, each multispectral image patch is vectorized before dimensionality reduction. For every one-vs-one classification task, PCA and feature normalization are fitted exclusively on the corresponding training partition. The resulting transformations are then applied unchanged to the validation, binary test, and multiclass global-test samples for that classifier. Consequently, validation and test images never contribute to preprocessing, model selection, or model fitting, ensuring a leakage-free evaluation throughout both the binary and multiclass experiments.

The final selected representation uses 32 PCA components. The selection process is described in Section~\ref{sec:selection}. PCA-64 is also evaluated as a sensitivity check but is not used for the main configuration because it does not significantly improve validation accuracy for the target nonlinear models under the final circuit.

\subsection{\texorpdfstring{Model Configurations and Hyperparameters}{Classical and Quantum Models}}

All PCA-based models receive the same pair-specific preprocessed feature vectors and are evaluated under the same fixed splits. The selected baselines separate linear classifiers, a strong nonlinear classical kernel, and a compact neural network whose parameter count is comparable to that of the VQC. The two quantum readouts share the same trained circuit: the VQC uses local Pauli-\(Z\) expectations and a linear head, while SVM-QK freezes the trained circuit and replaces the readout by a fidelity-kernel SVM. Table~\ref{tab:model_settings} summarizes the fixed hyperparameter choices. The ResNet-18 rows are included as direct image-based references and are not used for PCA or circuit selection.

\begin{table}[!h]

\centering
\caption{Fixed preprocessing, model, and training hyperparameters used throughout the experiments.}
\label{tab:model_settings}
\scriptsize
\setlength{\tabcolsep}{2pt}
\begin{tabular}{p{0.28\linewidth} p{0.62\linewidth}}
\hline
Component & Setting \\
\hline
Task protocol & EuroSAT-MS one-vs-one tasks; 45 class pairs; fixed 1400/300/300 train/validation/test samples per class; five training seeds. \\
PCA & Pair-specific PCA; fitted on the pairwise training partition only; main setting \(d=32\); PCA-64 used only as sensitivity check. \\
Normalization & Pair-specific normalization fitted on the training partition only and reused unchanged for validation, binary test, and global-test predictions. \\
Logistic regression & \(\ell_2\) logistic regression; \(C=1\); lbfgs; max iterations 5000. \\
SVM-linear & Linear-kernel SVM; \(C=1\). \\
SVM-RBF & RBF-kernel SVM; \(C=1\); \(\gamma=\texttt{scale}\). \\
Parameter-matched NN & MLP \(32\rightarrow8\rightarrow1\); ReLU; softplus loss; Adam; learning rate \(10^{-2}\); batch size 32; 500 epochs; validation checkpoint; 273 parameters. \\
VQC & \(q=4\), \(L=10\), alternating topology, offset 8; local \(Z\) readout with linear head; Adam; learning rate \(10^{-2}\); batch size 32/64; max 80 epochs; patience 40; 253 parameters. \\
SVM-QK (trained) & Frozen trained VQC circuit; fidelity kernel; \(\texttt{kernel=precomputed}\); \(C=1\). \\
ResNet-18 RGB & Direct image model; Sentinel-2 B04--B03--B02; ImageNet initialization; cross-entropy; Adam; learning rate \(10^{-3}\); batch size 64; max 50 epochs; patience 10. \\
ResNet-18 MS & Direct image model; all 13 EuroSAT-MS bands; first convolution expanded from ImageNet weights; otherwise same training settings as RGB. \\
\hline
\end{tabular}
\end{table}


\subsection{\texorpdfstring{Quantum Circuit Architecture}{Circuit Architecture}}

The selected circuit uses \(q=4\) qubits, \(L=10\) data re-uploading blocks,  each followed by an entanglement layer. It follows the data re-uploading idea of repeatedly injecting classical features into a trainable circuit \cite{PerezSalinas2020DataReuploading}, with the particular depth and topology selected by the validation ablation in Section~\ref{sec:selection}. Since the choice of encoding can strongly affect VQC expressivity \cite{schuld2021data_encoding_expressive_power}, we describe the actual angles/operations used by the circuit rather than treating the ansatz as a black box.

The circuit begins from \(\ket{0}^{\otimes q}\) and applies a parameter-only initialization block with one \(R_y\) and one \(R_z\) rotation per qubit. Each subsequent re-uploading block applies three input-dependent rotations to each qubit, followed by a fixed CNOT entangling layer. For block \(\ell\in\{0,\ldots,L-1\}\), qubit \(r\in\{0,\ldots,q-1\}\), PCA dimension \(d\), and offset \(s\), the two feature indices used on that qubit are
\begin{equation}
i_{1,\ell r}=(\ell q+r)\bmod d,
\qquad
i_{2,\ell r}=(\ell q+r+s)\bmod d .
\end{equation}

For the finally selected PCA-32 circuit, the validation study selected \(d=32\), \(q=4\), and \(s=8\). Thus the first index advances by four PCA components from one block to the next, while the offset selects a second component eight positions ahead. Across the ten blocks, every PCA component from 0 to 31 appears directly in at least one encoding angle. Thus, every PCA component contributes directly to the quantum feature map.

\begin{figure*}[!t]

\centering
\includegraphics[width=\textwidth]{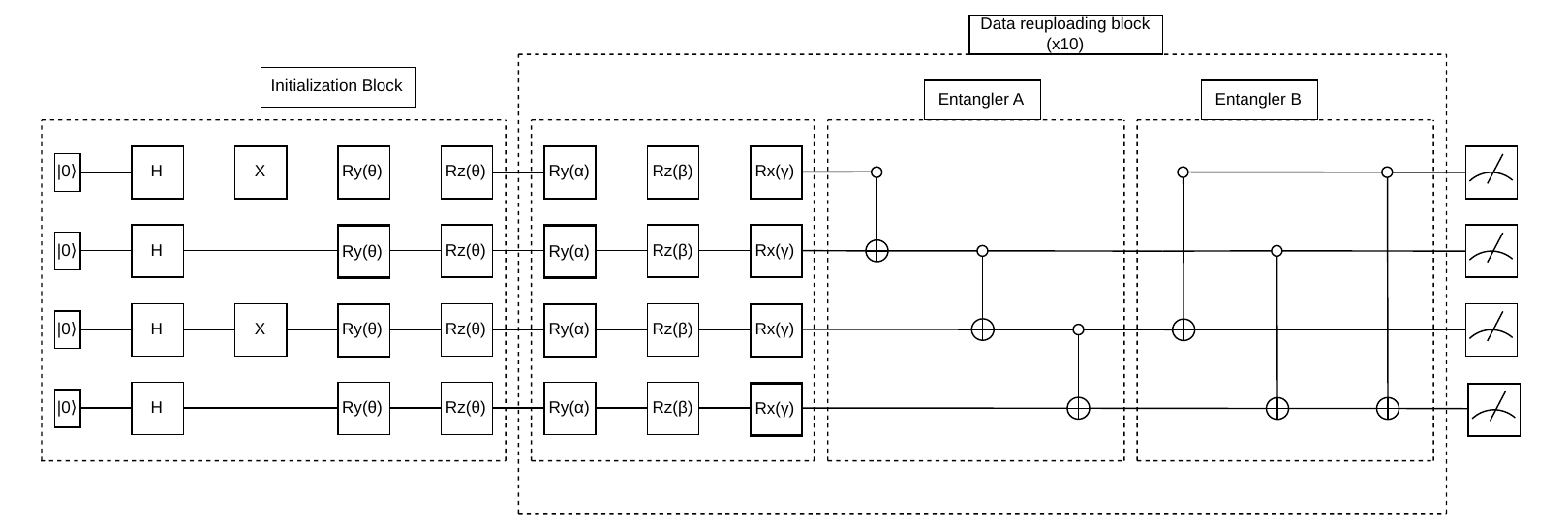}
\caption{Schematic of the selected four-qubit data re-uploading circuit. The final configuration uses PCA-32 inputs, \(L=10\) re-uploading blocks, offset 8, and alternating entanglement (between entangler A and B). The labels \(R_y(\alpha)\), \(R_z(\beta)\), and \(R_x(\gamma)\) are shorthand for block- and qubit-specific affine functions of selected PCA features with trainable coefficients. The same trained circuit is used either with local Pauli-\(Z\) measurements and a linear head or as a frozen feature map for the trained quantum-kernel SVM.}
\label{fig:circuit_architecture}
\end{figure*}

Let \(x_{i_{1,\ell r}}\) and \(x_{i_{2,\ell r}}\) denote the two selected preprocessed PCA features. The three data-dependent angles are affine functions of the selected PCA features with trainable coefficients:
\begin{align}
\alpha_{\ell r}(\boldsymbol{x};\boldsymbol{\theta})
&=
\pi\left(a_{\ell r}x_{i_{1,\ell r}}+b_{\ell r}\right),\\
\beta_{\ell r}(\boldsymbol{x};\boldsymbol{\theta})
&=
\pi\left(c_{\ell r}x_{i_{2,\ell r}}+d_{\ell r}\right),\\
\gamma_{\ell r}(\boldsymbol{x};\boldsymbol{\theta})
&=
\pi\left(e_{\ell r}\left(x_{i_{1,\ell r}}+x_{i_{2,\ell r}}\right)+f_{\ell r}\right).
\end{align}
The schematic gates \(R_y(\alpha)\), \(R_z(\beta)\), and \(R_x(\gamma)\) in Fig.~\ref{fig:circuit_architecture} therefore denote qubit- and block-specific rotations \(R_y(\alpha_{\ell r})\), \(R_z(\beta_{\ell r})\), and \(R_x(\gamma_{\ell r})\), not three shared global parameters. The trainable quantum parameter count is \(2q+6qL=248\); adding the \(q+1\) coefficients of the linear VQC head gives 253 trainable parameters.

Two CNOT layouts are compared, motivated by the known role of entangling structure in PQC expressibility \cite{Sim2019Expressibility}. The ring topology applies the cyclic nearest-neighbor pattern \((0,1),(1,2),(2,3),(3,0)\) in every block. The alternating topology uses that same ring pattern in even blocks and the cross-coupling pattern \((0,2),(1,3),(0,3)\) in odd blocks. For \(L=10\), the selected alternating circuit therefore uses \(5\times4+5\times3=35\) CNOT gates, compared with 40 CNOT gates for the \(L=10\) ring circuit. The comparison is consequently both an accuracy comparison and a two-qubit-gate cost comparison, since CNOT gates are typically the dominant source of hardware error.

\subsection{\texorpdfstring{Statistical Analysis}{Statistical Tests and Runtime Measurements}}

Accuracy is averaged over the five training seeds for each class pair, and class-pair means are used as the unit of analysis for paired model comparisons. This gives 45 paired units for EuroSAT-MS, 136 for So2Sat-LCZ42, and 15 for SAT-6. Unless otherwise stated, statistical comparisons use paired Wilcoxon signed-rank tests over class-pair means. When several architecture settings or model comparisons are considered as a family, Holm correction is applied within that family.

Using class-pair means rather than all 225 pair-seed values as independent samples avoids treating repeated training runs for the same class pair as independent observations. It also prevents class pairs with particularly low seed variance from disproportionately influencing the statistical tests. In the architecture ablation, paired comparisons are performed over the 20 representative class-pair means, whereas the final binary evaluation uses all 45 class-pair means. Accordingly, the reported \(p\)-values quantify the consistency of performance differences across class pairs rather than statistical significance at the image level.

Runtime and peak memory are recorded at the job level. The reported SVM-QK runtime includes kernel construction and SVM fitting after the corresponding VQC has already been trained. Consequently, the trained-kernel model inherits the computational cost of VQC training and should not be interpreted as an independent training runtime.

\subsection{\texorpdfstring{Implementation and Hardware}{Implementation and Hardware}}

All reported quantum simulations are exact noiseless state-vector simulations executed on CPU hardware. The experiments were run on an MSI Cyborg 15 A12VE laptop with a 12th Gen Intel Core i7-12650H processor, 10 physical cores, 16 logical processors, and 16 GB RAM, under Windows 11 Home. The reported experiments were performed without CUDA acceleration. The main pairwise PCA-32 and PCA-64 runs used eight process-level workers, with BLAS thread variables set to one thread per worker to avoid oversubscription. The recorded software environment used Python 3.10, PyTorch 2.7.1 CPU, scikit-learn 1.7.2, NumPy 2.2.6, and SciPy 1.15.3.

\FloatBarrier

\section{\texorpdfstring{Validation-Based Configuration Selection}{Validation-Based Configuration Selection}}

\label{sec:selection}

The final configuration is selected without using test accuracy. Because PCA dimensionality and circuit capacity interact through the data re-uploading schedule, we treat model selection as a configuration search rather than as two independent choices. Candidate configurations are compared using validation accuracy together with measures of feasibility and hardware relevance, including explained variance, feature coverage, trainable-parameter count, two-qubit gate count, and runtime. Held-out test results are reported only after the configuration is fixed, in Section~\ref{sec:results}.

\subsection{\texorpdfstring{Representative Pair Selection}{Representative Pair Selection}}

Running every candidate configuration over all 45 pairs and five seeds would be expensive because each setting requires many VQC trainings. For the configuration search and qubit-count studies, we therefore use a  representative subset of 20 pairs. The subset is selected using only validation metrics from pre-existing classical baselines; no quantum results and no test accuracies are used.

For each of the 45 pairs, validation accuracy from logistic regression, SVM-linear, SVM-RBF, and NN is converted to a percentile rank, and the four percentiles are averaged to form a composite difficulty score. The resulting difficulty scores partition the 45 class pairs into five quintiles, ranging from the hardest to the easiest. Four pairs are selected from each quintile, and the selection is constrained so that every EuroSAT class appears in exactly four selected pairs. The resulting subset therefore spans the full range of classification difficulty while maintaining uniform class coverage, reducing the risk that configuration selection is biased toward either particularly easy or particularly difficult class pairs.

\subsection{\texorpdfstring{PCA and Circuit Search}{PCA and Circuit Search}}

The configuration search proceeds in two stages. First, we examine the effect of PCA dimensionality while keeping the original compact circuit fixed. This identifies the range of PCA dimensions worth exploring. Second, because PCA dimensionality and circuit capacity interact through the data re-uploading schedule, the remaining candidate architectures are evaluated jointly using the representative subset introduced above.

Table~\ref{tab:pca_sensitivity} summarizes the initial PCA sensitivity analysis over 8, 16, and 32 components. PCA-8 is discarded because it consistently removes useful spectral information. Increasing the representation from PCA-8 to PCA-16 significantly improves validation accuracy for all nonlinear models, with Holm-adjusted paired Wilcoxon \(p\)-values of 0.004 for VQC, 0.003 for SVM-QK, 0.012 for NN, and 0.001 for SVM-RBF. The mean cumulative explained variance also increases from 71.85\% to 77.97\%.

\begin{table}[!h]
\centering
\caption{PCA sensitivity analysis. Validation accuracies are averaged over class-pair means. The PCA-64 row reports the final sensitivity check performed under the selected L=10 circuit, whereas the remaining rows correspond to the initial PCA sweep at fixed circuit depth. The NN column refers to the compact MLP used during the PCA search.}
\label{tab:pca_sensitivity}
\footnotesize
\begin{tabular}{c c c c c c}
\hline
PCA & Expl. var. & VQC & SVM-QK & NN & SVM-RBF \\
\hline
8  & 71.85 & 95.07 & 94.74 & 95.00 & 94.52 \\
16 & 77.97 & 95.60 & 95.31 & 95.43 & 95.53 \\
32 & 83.59 & 95.75 & 95.45 & 95.79 & \textbf{96.08} \\
64 & 88.45 & 96.26 & 96.07 & 95.88 & 96.08 \\
\hline
\end{tabular}
\end{table}

The comparison between PCA-16 and PCA-32 is intentionally performed at the same circuit depth (\(L=6\)) to isolate the effect of the retained PCA dimension. Under this controlled comparison, PCA-32 increases the mean explained variance to 83.59\% and slightly improves the validation accuracy of all nonlinear models, but none of these improvements remains statistically significant after Holm correction for VQC (\(+0.15\) percentage points, \(p_{\mathrm{Holm}}=1.000\)), SVM-QK (\(+0.15\), \(p_{\mathrm{Holm}}=1.000\)), or NN (\(+0.36\), \(p_{\mathrm{Holm}}=0.838\)). The largest improvement is observed for SVM-RBF (\(+0.55\) percentage points), but it is only borderline after correction (\(p_{\mathrm{Holm}}=0.057\)). Therefore, PCA-32 is not selected on the basis of dimensionality alone. Instead, it is carried forward into the joint architecture search together with the larger circuit configurations.

\begin{figure*}[!t]
\centering
\includegraphics[width=0.92\textwidth]{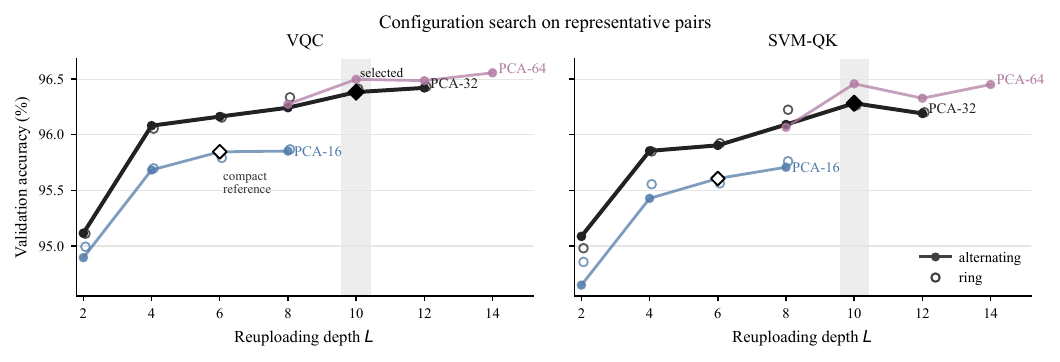}
\caption{Configuration-search trends on the 20 representative pairs. Solid curves correspond to the alternating entanglement topology for each PCA dimension, while hollow markers indicate the ring topology where evaluated. The compact reference configuration is PCA-16 with $L=6$ and alternating entanglement; the selected configuration is PCA-32 with $L=10$ and alternating entanglement.}
\label{fig:config_search}
\end{figure*}

Figure~\ref{fig:config_search} summarizes the joint configuration search, while Table~\ref{tab:config_search} reports the principal candidate architectures. The complete configuration grid is provided in Appendix~\ref{app:config_grid}. The compact reference configuration is PCA-16 with \(L=6\) and alternating entanglement. Relative to this reference, the configuration PCA-32 with \(L=10\) and alternating entanglement produces statistically significant improvements on the representative subset for both VQC (\(+0.54\) percentage points, \(p_{\mathrm{Holm}}=0.019\)) and SVM-QK (\(+0.68\) percentage points, \(p_{\mathrm{Holm}}=0.009\)). This comparison shows that increasing the retained PCA dimension together with the circuit capacity yields a meaningful improvement over the original compact architecture while remaining within the same four-qubit design.

\begin{table}[!h]

\centering
\caption{Key validation configurations on the 20 representative pairs. Validation accuracies (\%) are averaged over class-pair means and five training seeds.}
\label{tab:config_search}
\scriptsize
\setlength{\tabcolsep}{1.4pt}
\begin{tabular}{p{0.33\linewidth} p{0.29\linewidth} c c c}
\hline
Setting & Role & CNOTs & VQC & SVM-QK \\
\hline
PCA-16, \(L=6\), alt. & Compact reference & 21 & 95.85 & 95.61 \\
PCA-32, \(L=10\), alt. & Selected & 35 & 96.38 & 96.28 \\
PCA-32, \(L=12\), ring & Best PCA-32 VQC mean & 48 & 96.43 & 96.20 \\
PCA-64, \(L=14\), alt. & Best PCA-64 VQC mean & 49 & 96.56 & 96.45 \\
\hline
\end{tabular}
\end{table}

Having identified PCA-32 with \(L=10\) as the best-performing compact architecture, we next examine whether additional circuit capacity provides further benefit. Within the PCA-32 family, increasing the depth from \(L=10\) to \(L=12\) yields only marginal improvements: \(+0.04\) percentage points for the alternating topology (\(p_{\mathrm{Holm}}=1.000\)) and \(+0.05\) percentage points for the ring topology (\(p_{\mathrm{Holm}}=0.765\)) relative to the selected \(L=10\) alternating circuit. Likewise, PCA-64 is evaluated using deeper alternating circuits and offset 32 so that all 64 components participate in the data re-uploading schedule. Although PCA-64 produces slightly higher representative-subset means, its best evaluated configuration (PCA-64, \(L=14\), alternating) is not significantly better than the selected PCA-32, \(L=10\), alternating configuration for either VQC (\(+0.17\) percentage points, \(p=0.207\)) or SVM-QK (\(+0.17\), \(p=0.165\)). A final validation over all 45 class pairs also shows no significant advantage of PCA-64 over PCA-32 under the selected \(L=10\) circuit for VQC (\(+0.08\) percentage points, \(p=0.809\)), SVM-QK (\(+0.06\), \(p=0.732\)), NN (\(-0.10\), \(p=0.185\)), or SVM-RBF (\(-0.02\), \(p=0.928\)).

The entangling topology is selected using the same accuracy--cost criterion. The ring topology is not significantly more accurate than the alternating topology at \(L=10\) (\(+0.03\) percentage points, \(p_{\mathrm{Holm}}=1.000\)), while requiring 40 CNOT gates instead of 35. We therefore select the configuration PCA-32 with \(L=10\), offset 8, and alternating entanglement. This configuration significantly improves upon the original compact reference, explicitly encodes all 32 retained PCA components, lies near the validation-performance plateau, and avoids the additional two-qubit gate cost of the ring alternative.

\begin{figure}[!t]

\centering
\includegraphics[width=\linewidth]{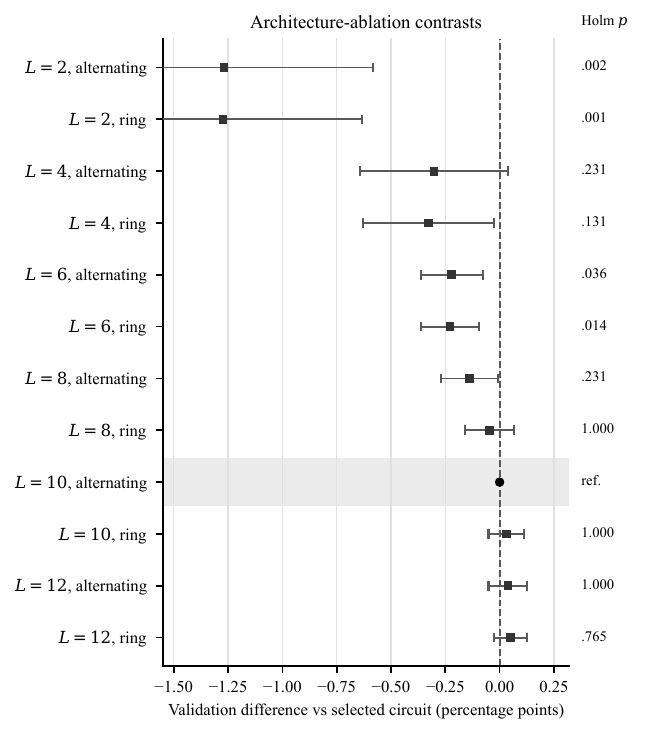}
\caption{Paired comparisons of the PCA-32 architectures based on VQC validation accuracy over the 20 representative pairs. Differences are measured against the selected \(L=10\) alternating circuit; horizontal intervals show 95\% confidence intervals over pairwise differences and the right column reports Holm-adjusted Wilcoxon \(p\)-values.}
\label{fig:architecture_ablation}
\end{figure}

\FloatBarrier

\section{Results}

\label{sec:results}

\subsection{\texorpdfstring{Binary One-vs-One Classification}{Binary One-vs-One Classification}}

Table~\ref{tab:pairwise_main} reports the final PCA-32 results over all 45 class pairs and five training seeds. The VQC substantially outperforms both logistic regression and the linear SVM, indicating that the learned PQC representation provides a useful nonlinear transformation of the PCA features. The parameter-matched NN is close to the VQC. The VQC is higher on validation by 0.26 percentage points (\(p=0.012\)), while the NN is higher on held-out test accuracy by 0.06 percentage points; the test difference is not significant (\(p=0.434\)).

Replacing the linear VQC readout with the trained SVM-QK improves the held-out test accuracy by $0.41$ percentage points (\(p=5.5\times10^{-8}\)). It also improves over SVM-linear by 2.50 percentage points (\(p=4.5\times10^{-8}\)) and over the parameter-matched NN by 0.35 percentage points (\(p=0.0011\)). The RBF-SVM remains higher than SVM-QK by 0.58 percentage points on the held-out test set (\(p=0.011\)), so the quantum-kernel result is best interpreted as narrowing the gap to a strong classical kernel rather than surpassing it.

Taken together, the pairwise results support a representation-level conclusion. The learned PQC feature map consistently outperforms the linear baselines, and the the fidelity-kernel readout extracts additional predictive information from the learned quantum representation. The remaining gap to SVM-RBF keeps the claim deliberately narrow: the trained states contain useful class structure, but the practical value of that structure depends on how it is read out and how much the readout costs.

\begin{table}[!t]

\centering
\caption{Final PCA-32 one-vs-one accuracy over all 45 class pairs. Values are means over class-pair means.}
\label{tab:pairwise_main}
\scriptsize
\setlength{\tabcolsep}{1.5pt}
\begin{tabular}{@{}p{0.52\linewidth} c c@{}}
\hline
Model & Validation & Test \\
\hline
Logistic regression & 92.77 & 93.20 \\
SVM-linear          & 92.84 & 93.32 \\
Parameter-matched NN & 95.93 & 95.47 \\
VQC                 & \textbf{96.18} & 95.41 \\
SVM-QK (trained)    & 96.01 & 95.82 \\
SVM-RBF             & 96.10 & \textbf{96.40} \\
\hline
\end{tabular}
\end{table}

\begin{figure}[!t]

\centering
\includegraphics[width=\linewidth]{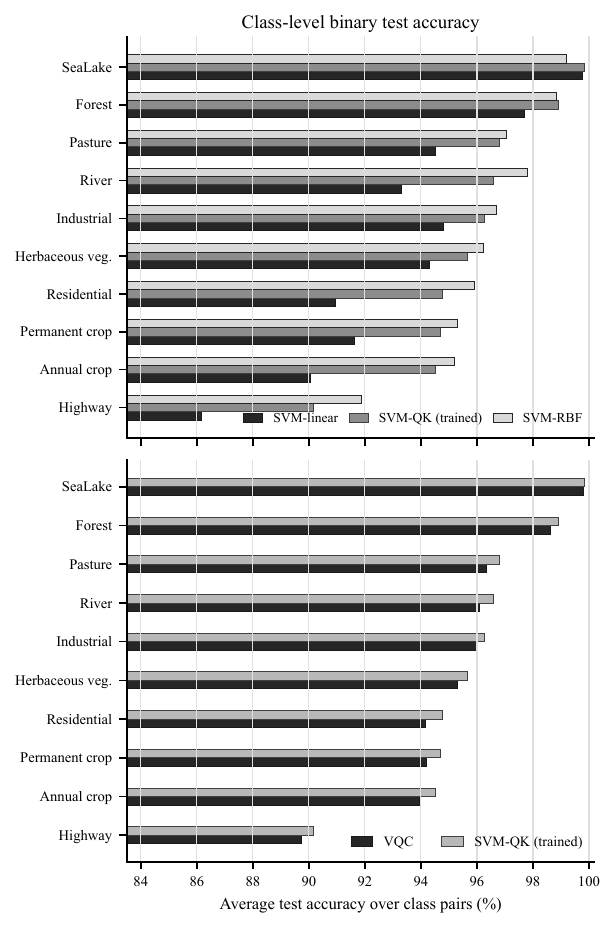}
\caption{Mean per-class binary test accuracy for the final PCA-32 setting. For each class, accuracy is averaged over the nine one-vs-one pairs containing that class and over five seeds. The top panel compares SVM-linear, SVM-QK (trained), and SVM-RBF; the bottom panel compares the linear VQC readout with the trained quantum-kernel readout derived from the same PQC.}
\label{fig:class_level_accuracy}
\end{figure}

Figure~\ref{fig:class_level_accuracy} summarizes the same binary test results by EuroSAT class. Each class value is the average over the nine pairwise tasks involving that class and over the five training seeds. The class-level analysis shows that the ordering is not driven by a single class: SVM-QK is consistently above the linear SVM and usually above the VQC readout, while SVM-RBF remains the strongest PCA-based reference for most classes.

\subsection{\texorpdfstring{Training-Induced Changes in the Quantum Kernel}{Trained Versus Untrained Quantum Kernels}}

The performance of the quantum-kernel readout depends critically on training the underlying PQC. Table~\ref{tab:trained_untrained} compares the untrained and trained fidelity kernels. Training increases the SVM-QK test accuracy from 89.73\% to 95.82\%. To understand why, we examine three diagnostics that characterize the geometry of the induced kernel. We report the mean fidelity separation
\begin{equation}
\Delta_F
=
\overline{K}_{\mathrm{same}}
-
\overline{K}_{\mathrm{different}},
\end{equation}
where \(\overline{K}_{\mathrm{same}}\) is the mean within-class fidelity and \(\overline{K}_{\mathrm{different}}\) is the mean between-class fidelity. Training increases \(\Delta_F\) from 0.019 to 0.290, indicating that same-class states become much more similar relative to different-class states.

We also report centered kernel-target alignment,
\begin{equation}
\mathrm{KTA}
=
\frac{\langle K_c,Y_c\rangle_F}
{\|K_c\|_F\|Y_c\|_F},
\end{equation}
where \(K_c\) is the centered Gram matrix and \(Y_c\) is the centered label kernel. This is the cosine similarity between the kernel similarities and the same-class/different-class label structure after centering. It increases from 0.152 to 0.656 after training. Finally, the participation-ratio effective rank,
\begin{equation}
d_{\mathrm{eff}}
=
\frac{\left(\sum_i \lambda_i\right)^2}
{\sum_i \lambda_i^2},
\end{equation}
increases from 1.75 to 14.77, where \(\lambda_i\) are the eigenvalues of the Gram matrix. This suggests that training distributes information across a richer set of directions in the learned feature space than the untrained kernel. The practical impact of these geometric changes is most evident on the hardest class pairs. On the hardest quintile of class pairs, training improves validation accuracy by 8.46 percentage points and test accuracy by 7.03 percentage points, with all nine pairwise differences favoring the trained kernel.

\begin{table}[!t]

\centering
\caption{Trained versus untrained quantum-kernel diagnostics for PCA-32. Values are means over the 45 class-pair means.}
\label{tab:trained_untrained}
\footnotesize
\begin{tabular}{l c c}
\hline
Metric & Untrained QK & Trained QK \\
\hline
Validation accuracy (\%) & 89.57 & \textbf{96.01} \\
Test accuracy (\%)       & 89.73 & \textbf{95.82} \\
Same-minus-different fidelity & 0.019 & \textbf{0.290} \\
Centered kernel-target alignment & 0.152 & \textbf{0.656} \\
Effective rank, participation ratio & 1.75 & \textbf{14.77} \\
Effective rank, spectral entropy & 3.96 & \textbf{52.51} \\
\hline
\end{tabular}
\end{table}

Together, these diagnostics show that the observed improvement cannot be attributed solely to the SVM optimization stage. Training changes the induced similarity matrix: same-class and different-class fidelities separate more strongly, the centered Gram matrix becomes more aligned with the label kernel, and the effective rank increases. This provides empirical support for reusing the trained PQC as a kernel feature map, while still leaving the final comparison to held-out accuracy.

\subsection{\texorpdfstring{Multiclass One-vs-One Voting}{Multiclass One-vs-One Voting}}

The trained pairwise classifiers are also evaluated are also evaluated in a full 10-class one-vs-one voting scheme on the common global test partition. Table~\ref{tab:ovo_multiclass} reports accuracy and macro-F1. The multiclass accuracies are lower than the mean binary accuracies because each 10-class prediction depends on 45 binary decisions, and errors on visually or spectrally similar classes can change the final vote.

The multiclass evaluation preserves the ordering observed in the pairwise experiments. VQC improves strongly over the linear baselines and is slightly higher than the parameter-matched NN. SVM-QK achieves the best quantum/PQC-based result, with 80.63\% accuracy and 80.71\% macro-F1. SVM-RBF remains strongest among the PCA-based models, reaching 82.73\% accuracy. Thus, the trained-kernel readout improves the PQC-based multiclass result, but it does not remove the gap to the classical RBF kernel.

\begin{table}[!t]

\centering
\caption{PCA-32 multiclass one-vs-one voting results on the global test set. Values are averages  over five training seeds where applicable.}
\label{tab:ovo_multiclass}
\footnotesize
\begin{tabular}{l c c}
\hline
Model & Accuracy & Macro-F1 \\
\hline
Logistic regression & 69.30 & 68.77 \\
SVM-linear          & 69.87 & 69.30 \\
Parameter-matched NN & 79.51 & 79.31 \\
VQC                 & 79.98 & 79.77 \\
SVM-QK (trained)    & 80.63 & 80.71 \\
SVM-RBF             & \textbf{82.73} & \textbf{82.68} \\
\hline
\end{tabular}
\end{table}

\FloatBarrier

\subsection{\texorpdfstring{Qubit-Count Sweep}{Qubit-Count Sweep}}

\label{subsec:qubit_sweep}

The selected four-qubit circuit is compact, so we also examine what happens when the number of qubits is increased while keeping the PCA-32 inputs, \(L=10\), offset-based reuploading, and alternating entanglement. This experiment is not part of the configuration-selection procedure; rather, it is an exploratory study  on the same 20 representative pairs, of how the architecture scales with the number of qubits. For \(q>4\), the circuit uses the same rule for distributing PCA components across qubits and reuploading blocks. This increases the Hilbert-space dimension and the number of trainable parameters, but it does not introduce a separately tuned architecture for each qubit count.

\begin{figure}[!h]

\centering
\includegraphics[width=\linewidth]{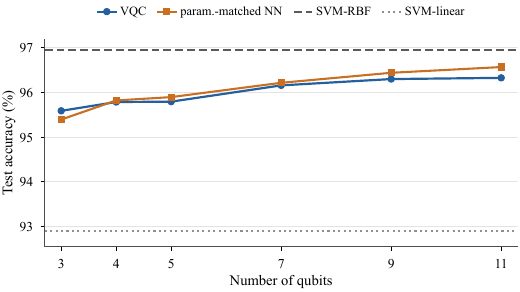}
\caption{Exploratory qubit-count sweep on the 20 representative pairs. Mean VQC accuracy improves with qubit count before gradually saturating. Parameter-matched NNs follow a similar trend and slightly exceed the VQC at larger \(q\).}
\label{fig:qubit_vqc_classical}
\end{figure}

Figure~\ref{fig:qubit_vqc_classical} compares VQC with SVM-linear, SVM-RBF, and parameter-matched NNs. VQC accuracy increases from 95.59\% at \(q=3\) to 96.32\% at \(q=11\), with most of the gain occurring at lower qubit counts. The parameter-matched NN follows a similar trend and slightly exceeds VQC at larger \(q\), reaching 96.57\% at \(q=11\). Both remain below the SVM-RBF reference on the representative subset.

\FloatBarrier

\begin{figure}[!h]

\centering
\includegraphics[width=\linewidth]{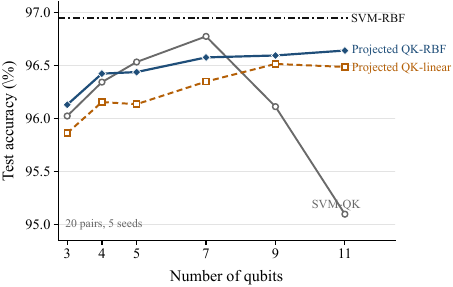}
\caption{Kernel readouts in the qubit-count sweep. The global fidelity SVM-QK improves up to \(q=7\) and then degrades, while projected kernels based on local and two-qubit Pauli observables remain comparatively stable as \(q\) increases.}
\label{fig:qubit_kernel_readouts}
\end{figure}

The fidelity-kernel readout behaves differently, as shown in Fig.~\ref{fig:qubit_kernel_readouts}. SVM-QK improves up to \(q=7\), reaching 96.77\%, but then drops sharply at \(q=9\) and \(q=11\). This behavior is consistent with kernel concentration: as the state space grows, global fidelities can become less informative unless the embedding and readout are adapted. To better understand this behavior, we also evaluate a projected quantum kernel. Instead of using global state fidelities, the projected kernel is constructed from local \(X,Y,Z\) expectations and two-qubit Pauli expectations on qubit pairs directly coupled by the circuit entanglers, then applies a classical kernel to the resulting projected feature vectors. We evaluate both a linear outer kernel and an RBF outer kernel; the linear projected kernel is already stable at larger \(q\), while the RBF projected kernel gives the strongest exploratory result, reaching 96.64\% at \(q=11\). The experiment is exploratory, but it suggests that local or projected quantum representations may be more robust than global fidelity kernels in larger state spaces.

The projected-kernel experiment helps clarify the limitations of the global fidelity kernel at larger qubit counts. Increasing the number of qubits expands the Hilbert space exponentially, but the number of trainable parameters and the number of training samples grow much more slowly. In such a regime, raw state fidelities may lose discriminative power for classification. Projecting the states onto local single-qubit and two-qubit Pauli expectation values  reduces the readout to observables that are more directly tied to local structure in the trained state. The result is not presented as an optimized new model, but it shows that the drop of SVM-QK at larger \(q\) is not necessarily a failure of the trained circuit alone; it is also a readout-design issue. These observations further support the central theme of the paper: the usefulness of a learned quantum representation depends not only on the representation itself but also on how it is interrogated.

\FloatBarrier

\subsection{\texorpdfstring{Additional Dataset Evaluation}{External Dataset Checks}}

\label{subsec:external_datasets}

The final PCA-32, \(L=10\) alternating circuit is also evaluated on So2Sat-LCZ42 and SAT-6 without any additional model selection. These experiments address a narrower question than the EuroSAT study: given the circuit and preprocessing protocol selected on EuroSAT, does the same readout behavior persist on datasets with different label semantics and difficulty?

Table~\ref{tab:so2sat_results} reports the So2Sat-LCZ42 results. The VQC achieves the highest mean validation accuracy, the highest mean pairwise test accuracy, 86.58\%, and the highest multiclass OvO accuracy, 42.46\%. On validation, the VQC is higher than the parameter-matched NN by \(+0.81\) percentage points (\(p_{\mathrm{Holm}}=2.7\times10^{-6}\)) and higher than SVM-RBF by \(+2.92\) percentage points (\(p_{\mathrm{Holm}}=7.1\times10^{-18}\)). On the held-out test pairs, the same ordering remains in the means, but the margins over the NN and SVM-RBF are not significant after correction: \(+0.48\) percentage points over the NN (\(p_{\mathrm{Holm}}=0.090\)) and \(+0.74\) percentage points over SVM-RBF (\(p_{\mathrm{Holm}}=0.425\)). The trained SVM-QK is essentially tied with SVM-RBF on mean pairwise test accuracy, 85.86\% versus 85.84\%, but it is lower than the VQC. The paired test comparison VQC minus SVM-QK is \(+0.73\) percentage points over the 136 class pairs and remains significant after Holm correction (\(p_{\mathrm{Holm}}=3.1\times10^{-5}\)). Thus, the improvement of the trained fidelity kernel over the VQC readout observed on EuroSAT does not transfer unchanged to LCZ classification.

\begin{table}[!h]

\centering
\caption{External check on So2Sat-LCZ42 using the fixed EuroSAT-selected PCA-32, \(L=10\) alternating circuit. Pairwise values are averages over the 136 class pairs. Multiclass values are averages over five training seeds. }
\label{tab:so2sat_results}
\scriptsize
\setlength{\tabcolsep}{2pt}
\begin{tabular}{l c c c c}
\hline
Model & Pair val. & Pair test & OvO acc. & OvO macro-F1 \\
\hline
Logistic regression & 83.21 & 83.61 & 28.68 & 29.33 \\
SVM-linear          & 83.11 & 83.34 & 29.15 & 29.46 \\
SVM-RBF             & 85.03 & 85.84 & 40.09 & 39.57 \\
Parameter-matched NN & 87.15 & 86.10 & 41.06 & 41.23 \\
VQC                 & \textbf{87.95} & \textbf{86.58} & \textbf{42.46} & \textbf{43.05} \\
SVM-QK (trained)    & 85.72 & 85.86 & 39.76 & 39.93 \\
\hline
\end{tabular}
\end{table}

The reversal on So2Sat-LCZ42 is consistent with the structure of the task. EuroSAT classes are mostly land-cover or material categories such as forest, water, crop, road, and residential areas. So2Sat-LCZ42 uses LCZ labels such as compact high-rise, open mid-rise, sparsely built, dense trees, and low plants. These labels primarily encode urban morphology, compactness, and building height rather than only spectral material. In our controlled run, only Sentinel-2 optical channels are used to match EuroSAT-MS; Sentinel-1 SAR channels, which may be informative for urban structure, are deliberately excluded. The So2Sat result should therefore be read as an external stress test of the selected spectral-PCA protocol, not as an optimized LCZ system.

Table~\ref{tab:external_geometry} gives the diagnostic evidence behind this interpretation. PCA-32 retains a similar amount of variance on So2Sat-LCZ42 as on EuroSAT-MS, so the readout reversal is not explained by lower total PCA variance. The divergence emerges after the learned quantum representation is formed. The full-state fidelity kernel is less aligned with the labels on So2Sat-LCZ42 than on EuroSAT-MS: the test kernel-target alignment drops from 0.640 to 0.493, the same-minus-different fidelity gap drops from 0.288 to 0.212, and fidelity AUC drops from 0.820 to 0.751. The measured \(Z\)-space representation is also less separated on So2Sat-LCZ42, but it remains sufficiently discriminative for the local VQC readout. These observations help explain why the same trained circuit can favor the fidelity-kernel readout on EuroSAT while favoring the local VQC readout on LCZ labels.

\begin{table}[!h]

\centering
\caption{Readout-geometry diagnostics on the EuroSAT-MS and So2Sat-LCZ42 test splits. Values are averages over the class pairs. \(Z\)-KTA denotes kernel-target alignment after applying an RBF kernel to the measured \(Z\)-expectation vectors. Full-QK KTA and fidelity AUC are computed from the trained state-fidelity kernel.}
\label{tab:external_geometry}
\scriptsize
\setlength{\tabcolsep}{2pt}
\begin{tabular}{l c c c c c}
\hline
Dataset & PCA var. & \(Z\)-KTA & Full-QK KTA & Fid. gap & QK--VQC \\
\hline
EuroSAT-MS & 83.59 & 0.798 & 0.640 & 0.288 & \(+0.41\) pp \\
So2Sat-LCZ42 & 85.22 & 0.579 & 0.493 & 0.212 & \(-0.73\) pp \\
\hline
\end{tabular}
\end{table}

SAT-6 gives a different kind of external check, shown in Table~\ref{tab:sat6_results}. It is closer to EuroSAT in label type because the classes are land-cover categories, but it is substantially easier under the present protocol. All nonlinear models exceed 98.8\% mean pairwise test accuracy, and many pairwise tasks are saturated. The trained SVM-QK gives the highest mean pairwise test accuracy, 99.00\%, and the highest OvO accuracy, 95.61\%, but the differences are small. Across the 15 class pairs, the overall Friedman test across the six models is not significant (\(p=0.171\)). The SVM-QK advantage over VQC is \(+0.14\) percentage points on pairwise test accuracy, with unadjusted \(p=0.035\) but Holm-adjusted \(p=0.244\). Its difference from SVM-RBF is only \(+0.05\) percentage points (\(p=0.553\)). SAT-6 therefore supports competitiveness of the trained quantum kernel on a land-cover dataset, but it is too saturated to support a strong superiority claim.

\begin{table}[!h]

\centering
\caption{External check on SAT-6 using the fixed EuroSAT-selected PCA-32, \(L=10\) alternating circuit. Pairwise values are averages over the 15 class pairs. Multiclass values are averages over five training seeds.}
\label{tab:sat6_results}
\scriptsize
\setlength{\tabcolsep}{2pt}
\begin{tabular}{l c c c c}
\hline
Model & Pair val. & Pair test & OvO acc. & OvO macro-F1 \\
\hline
Logistic regression & 97.71 & 97.71 & 89.17 & 89.21 \\
SVM-linear          & 97.96 & 97.89 & 89.94 & 89.98 \\
SVM-RBF             & 98.97 & 98.96 & 95.28 & 95.29 \\
Parameter-matched NN & 99.07 & 98.83 & 94.57 & 94.58 \\
VQC                 & \textbf{99.22} & 98.86 & 95.00 & 95.01 \\
SVM-QK (trained)    & 99.13 & \textbf{99.00} & \textbf{95.61} & \textbf{95.62} \\
\hline
\end{tabular}
\end{table}

Taken together, the external datasets sharpen rather than weaken the main conclusion. The trained PQC representation transfers across datasets, but the most effective readout is not universal. On EuroSAT-MS, the trained fidelity kernel extracts additional held-out signal from the same circuit. On SAT-6, the same effect is visible only as a small nonsignificant trend because the task is already near saturation. On So2Sat-LCZ42, the global fidelity geometry is less label-aligned, and the local VQC readout is stronger. The practical lesson is that quantum-kernel readouts should be evaluated through their induced geometry on the target data, not assumed to dominate local measurement heads by construction.

Together with the qubit-count study, these results reinforce the view that the effectiveness of a quantum readout depends on the geometry induced by the learned representation rather than on the representation alone.

\FloatBarrier

\subsection{\texorpdfstring{Image-Based ResNet Baselines}{Image-Based ResNet Baselines}}

The preceding experiments deliberately focus on compact PCA representations that discard spatial information, allowing different quantum and classical readouts to be compared under identical spectral inputs within the same controlled one-versus-one classification framework. An obvious complementary question is how this compact representation compares with a conventional image classifier that operates directly on the original multispectral images. To address this question, we evaluate ResNet-18 on the full EuroSAT-MS image patches. Unlike the PCA-based experiments, these models are trained directly for standard 10-class image classification rather than through the one-versus-one protocol. Consequently, the ResNet results should be interpreted as application-oriented reference baselines rather than controlled counterparts of the PCA-based study.

Two ResNet-18 variants are considered. The first uses the standard Sentinel-2 RGB composite (B04--B03--B02) with ImageNet initialization. The second operates on all 13 Sentinel-2 spectral bands. To accommodate the additional input channels, the first convolutional layer is expanded from three to thirteen channels by averaging the pretrained RGB filters across the channel dimension, replicating the resulting filters over all input bands, and scaling them by \(3/13\). For both models, channel means and standard deviations are estimated exclusively from the training split. Training uses cross-entropy loss with the Adam optimizer, learning rate \(10^{-3}\), batch size 64, evaluation batch size 128, a maximum of 50 epochs, and early stopping with a patience of 10 epochs. No data augmentation is applied.

Table~\ref{tab:resnet} summarizes the results. As expected, the multispectral ResNet outperforms its RGB counterpart, reaching 96.07\% test accuracy and 96.07\% macro-F1, approximately one percentage point higher than the RGB model. These accuracies are substantially higher than those obtained with the compact PCA protocol. This difference is expected because the ResNet models exploit the complete spatial and spectral content of the original image patches, whereas the PCA-based pipeline intentionally removes spatial information in order to isolate the role of compact spectral representations and their associated readouts.

A direct protocol-to-protocol comparison would require training separate image-based ResNet models for every binary one-versus-one task, followed by the same voting procedure used throughout the paper. Such an evaluation would greatly increase the computational cost while addressing a different question from the one considered here. The purpose of the present comparison is instead to provide a standard remote-sensing reference using a widely adopted CNN architecture trained under its natural multiclass formulation.

The ResNet baselines therefore complement, rather than compete with, the controlled PCA study. While the compact PCA protocol is designed to investigate learned spectral representations and the effect of alternative quantum readouts under identical inputs, the image-based ResNet models illustrate the performance attainable when conventional CNNs are allowed to exploit the full spatial and spectral information available in the original multispectral images. At present, this controlled PCA framework is also motivated by the computational cost of classically simulating parameterized quantum circuits at larger qubit counts. Extending the present study to image-level quantum models operating directly on high-dimensional multispectral inputs would require substantially larger quantum circuits than are currently practical to simulate. As larger quantum hardware and more scalable simulation techniques become available, revisiting this comparison with direct image-based quantum models will be an important direction for future work.

\begin{table}[!h]

\centering
\caption{Direct image-based ResNet-18 baselines on EuroSAT-MS. Runtime is reported as the average wall-clock time per training seed on CPU. Peak RAM is the maximum recorded resident-set size over the five seeds.}
\label{tab:resnet}
\scriptsize
\setlength{\tabcolsep}{2pt}
\begin{tabular}{l c c c c c}
\hline
Input & Val. & Test & F1 & Params & Time / RAM \\
\hline
RGB       & 95.11 & 95.07 & 95.06 & 11.18M & 91.25 m / 2.24 GB \\
13-band MS & \textbf{95.93} & \textbf{96.07} & \textbf{96.07} & 11.21M & 105.45 m / 2.71 GB \\
\hline
\end{tabular}
\end{table}

\FloatBarrier

\section{\texorpdfstring{Discussion}{Discussion}}

\label{sec:discussion}

\subsection{\texorpdfstring{Representation and Readout}{Representation and Readout}}

The central empirical finding is that the trained PQC is more informative than any single readout alone suggests. On EuroSAT-MS, the VQC improves over linear baselines, indicating that the circuit learns a nonlinear transformation of the PCA features. Reusing the same trained circuit as a fidelity kernel further improves performance on held-out binary tasks and in multiclass one-versus-one voting. The trained-versus-untrained diagnostics support this interpretation: training increases fidelity separation, kernel-target alignment, and effective rank.

The external datasets refine this conclusion. The trained quantum kernel is not universally the best readout of a trained PQC. While SVM-QK performs slightly better on SAT-6, the difference is not statistically significant after correction. On So2Sat-LCZ42, where the labels reflect local-climate-zone morphology rather than primarily spectral land-cover material, the VQC readout performs better than the fidelity kernel. These results suggest that a trained PQC may contain discriminative structure that different readouts exploit differently, and that the most effective readout depends on how the learned representation aligns with the target task.

The trained quantum kernel does not outperform SVM-RBF, which remains a strong nonlinear baseline on the same PCA inputs at substantially lower computational cost. Consequently, the present results should be interpreted as evidence that the PQC learns a meaningful representation rather than as a claim of quantum advantage. More broadly, they suggest that designing effective quantum machine learning models involves not only learning suitable quantum representations but also understanding which readouts best exploit them. Identifying the characteristics of datasets and learning problems that naturally favor particular quantum representations and readout strategies remains an interesting direction for future work.

\subsection{\texorpdfstring{Relation to Image-Based Baselines}{Relation to Image-Based Baselines}}

The ResNet-18 experiments provide an application-oriented reference rather than a direct counterpart to the PCA-based quantum models. Unlike the VQC and SVM experiments, which deliberately operate on compact PCA representations to isolate the effect of the learned quantum representation and its readout, the ResNet models exploit the full spatial and spectral content of the original image patches through deep convolutional architectures. Their superior performance is therefore expected and highlights the importance of comparing emerging QML methods with modern remote-sensing baselines when assessing practical image-classification performance.

At the same time, the comparison illustrates an important limitation of current QML studies. The PCA protocol is not only a controlled experimental setting but also a practical compromise imposed by the computational cost of PQcs. Simulating larger quantum systems during repeated optimization remains computationally prohibitive, making compact spectral representations the natural choice for systematic investigations of quantum feature maps and readout strategies.

As larger quantum processors and more scalable simulation techniques become available, an important future direction will be to move beyond compressed spectral vectors and develop quantum models that operate directly on the high-dimensional spectral and spatial structure of multispectral imagery. Such models could learn correlations across spectral bands without an intermediate PCA compression step, potentially allowing quantum representations to exploit richer spectral-spatial information while remaining faithful to the original image geometry.

\subsection{\texorpdfstring{Noise and Hardware Feasibility}{Noise and Hardware Feasibility}}\label{subsec:noise}

All quantum results in this work are obtained with exact noiseless state-vector simulation. They should therefore be interpreted as an idealized study of representation and readout quality, not as a prediction of current hardware performance. We do not expect the small noiseless gap between SVM-QK and the linear VQC head to survive automatically on unmitigated present-day hardware. Likewise, there is no evidence that the remaining gap between SVM-QK and SVM-RBF would be preserved under realistic device noise and finite-shot kernel estimation.

The selected four-qubit, \(L=10\) circuit contains 128 single-qubit rotations, 35 logical CNOT gates, and a logical layer count of 44 before hardware transpilation. Executing the circuit on hardware would additionally require device-specific routing, native-gate decomposition, while introducing idling, measurement noise, crosstalk, and calibration drift. Barren plateaus are a known obstacle in variational circuits \cite{mcclean2018barren}, and noise-induced barren-plateau behavior has been shown for variational quantum algorithms under local noise assumptions \cite{wang2021noise_induced}. This does not imply that the present four-qubit circuit exhibits a barren plateau, but it highlights a potential limitation as circuit depth increases.

Implementing the fidelity kernel is substantially more demanding than performing inference with the direct VQC readout. Training and evaluating embedding kernels on near-term devices requires careful treatment of finite shots, noise, and mitigation \cite{Hubregtsen,peters2021noisy_kernel}. An inversion-test estimate of
\(K_{\boldsymbol{\theta}}(\boldsymbol{x},\boldsymbol{x}')\)
implements approximately
\begin{equation}
U_{\boldsymbol{\theta}}^{\dagger}(\boldsymbol{x})
U_{\boldsymbol{\theta}}(\boldsymbol{x}'),
\end{equation}
and therefore effectively doubles, giving about 256 single-qubit rotations and 70 logical CNOT gates before routing and decomposition. A SWAP-test implementation would require additional qubits and controlled operations. Kernel estimation is therefore expected to be more sensitive to accumulated gate errors than local \(Z\)-measurement inference.

Although current superconducting processors continue to improve, two-qubit error rates and device connectivity remain nonuniform across couplings \cite{ibm2025heron}. Under a simplified global depolarizing model,
\begin{equation}
\widetilde{\rho}(\boldsymbol{x})
=
(1-\lambda)\rho(\boldsymbol{x})
+\lambda\frac{I}{2^q},
\end{equation}
the fidelity kernel becomes
\begin{equation}
\widetilde{K}(\boldsymbol{x},\boldsymbol{x}')
=
(1-\lambda)^2K(\boldsymbol{x},\boldsymbol{x}')
+\frac{2\lambda-\lambda^2}{2^q}.
\label{eq:noisy_kernel}
\end{equation}
Thus, the difference between within-class and between-class fidelities is contracted by \((1-\lambda)^2\). This contraction result is specific to unital noise such as depolarizing noise. Non-unital channels such as amplitude damping can push states toward biased fixed points, and coherent or correlated errors can distort different kernel entries in nonuniform ways.

Finite-shot estimation introduces an additional source of uncertainty. More generally, concentration effects can make quantum-kernel entries less informative as the effective Hilbert space grows \cite{thanasilp2024concentration}. If a fidelity is estimated from \(S\) Bernoulli measurements, its standard error obeys
\begin{equation}
\sigma_{\widehat{K}_{ij}}
=
\sqrt{\frac{K_{ij}(1-K_{ij})}{S}}
\leq
\frac{1}{2\sqrt{S}}.
\end{equation}
For 1024 shots, the worst-case standard error is approximately 1.56\%, decreasing to 0.50\% for 10\,000 shots. In the present binary protocol, each pair and seed requires \(N_{\mathrm{tr}}=2800\), \(N_{\mathrm{val}}=600\), and \(N_{\mathrm{te}}=600\). Exploiting symmetry in the training Gram matrix, one pair-seed requires
\begin{equation}
\frac{N_{\mathrm{tr}}(N_{\mathrm{tr}}+1)}{2}
+N_{\mathrm{val}}N_{\mathrm{tr}}
+N_{\mathrm{te}}N_{\mathrm{tr}}
=
7{,}281{,}400
\end{equation}
distinct kernel estimates. At 1024 shots per entry, this is approximately \(7.46\times10^9\) shots per pair-seed and \(1.68\times10^{12}\) shots over all 45 pairs and five seeds, excluding training circuits and calibration overhead. The complete training and kernel-evaluation protocol is therefore impractical on present hardware at its current scale.

Recent work on quantum-efficient kernel estimation, randomized measurements, projected quantum kernels, and shadow-based representations offers possible routes to reduce this burden \cite{peters2021noisy_kernel,coelho2025quantumefficient,haug2023randomized,huang2021power,jerbi2024shadows}. hese approaches approximate or replace the exact fidelity kernel used in the main experiments, so their effect on classification must be evaluated empirically. The projected-kernel sweep in Section~\ref{subsec:qubit_sweep} is one step in this direction.

Overall, the noiseless simulations should be viewed as an upper-bound study of representation quality under ideal state preparation, optimization, and kernel evaluation. They do not imply that the same protocol can be executed on current hardware with comparable performance. A practical implementation would require a substantially smaller kernel-evaluation budget together with shot-aware optimization, device-specific transpilation, and likely some combination of error mitigation or alternative local/projected readouts.

\subsection{\texorpdfstring{Simulation and Hardware Costs}{Computational Cost}}
\label{subsec:cost}

The simulation and hardware costs should be distinguished. In simulation, the VQC is trained using reverse-mode automatic differentiation through the state-vector representation. For fixed qubit count, circuit depth, and batch size, the computational cost grows approximately linearly with the number of training samples and training epochs, whereas the memory required for state-vector simulation scales as \(\mathcal{O}(2^q)\). By contrast, hardware training would typically rely on parameter-shift gradients, requiring additional shifted circuit evaluations for every trainable parameter and measured expectation value under finite-shot sampling \cite{schuld2019gradients}.

The trained SVM-QK inherits the cost of VQC training and additionally requires dense kernel construction. For \(N\) training samples and \(M\) evaluation samples, constructing the exact fidelity kernel requires \(\mathcal{O}(N^2)\) training overlaps and \(\mathcal{O}(MN)\) evaluation overlaps. In simulation, quantum states are cached after VQC training, allowing kernel overlaps to be computed efficiently by matrix multiplication and making kernel evaluation substantially cheaper than on quantum hardware for the four-qubit circuits considered here. In the final PCA-32 runs, the mean runtime per pair-seed was approximately 269 s for VQC training and evaluation, compared with 141 s for the parameter-matched NN, 0.24 s for SVM-RBF, and 0.17 s for SVM-linear. After VQC training, the subsequent kernel construction and classical SVM fit required an additional 1.6 s per pair-seed in the cached simulator. This overhead excludes the cost of training the underlying PQC and should not be interpreted as representative of hardware execution.

The measured wall-clock times illustrate the practical cost of the complete experimental pipeline (Table~\ref{tab:runtime_main}). The complete PCA-32/PCA-64 EuroSAT pipeline, including both quantum and classical stages together with multiclass one-versus-one aggregation, required 9.50 h on the local CPU system described in Section~\ref{sec:protocol}. The final PCA-32 quantum stage completed 225 pair-seed jobs in 3.22 h using eight workers. The separate parameter-matched NN run used the same global train-validation-test split and completed its 225 pair-seed jobs in 1.15 h. The external dataset experiments used the same CPU-only implementation together with the fixed configuration selected from the EuroSAT-MS study.

\begin{table*}[!t]

\centering
\caption{Recorded CPU wall times for the experiments reported in this work. Pair-seed means are measured inside completed jobs; stage wall times include worker scheduling and process overhead.}
\label{tab:runtime_main}
\footnotesize
\begin{tabular}{l c c c c}
\hline
Stage & Jobs / seeds & Workers & Wall time & Mean job-level time \\
\hline
PCA-32 VQC + trained SVM-QK & 225 pair-seeds & 8 & 3.22 h & VQC 268.6 s; QK extra 1.6 s \\
PCA-32 classical sweep & 225 pair-seeds & 8 & 1.71 h & compact NN 210.4 s; SVM-RBF 0.24 s \\
PCA-32 parameter-matched NN rerun & 225 pair-seeds & 8 & 1.15 h & NN training 140.7 s \\
PCA-32 multiclass OvO aggregation & 5 seeds & -- & 0.62 min & vote aggregation only \\
PCA-64 VQC + trained SVM-QK & 225 pair-seeds & 8 & 3.11 h & same circuit, offset 32 \\
PCA-64 classical sweep & 225 pair-seeds & 8 & 1.43 h & SVMs + compact NN \\
So2Sat-LCZ42 external check & 680 pair-seeds & 8 & 11.30 h & VQC 280.3 s; QK extra 1.5 s \\
SAT-6 external check & 75 pair-seeds & 8 & 37.3 min & VQC 132.4 s; QK extra 0.65 s \\
ResNet-18 RGB & 5 seeds & CPU & 7.62 h & 91.25 min / seed \\
ResNet-18 13-band MS & 5 seeds & CPU & 8.80 h & 105.45 min / seed \\
\hline
\end{tabular}
\end{table*}

The largest recorded per-job resident-set peak for the PCA-32 quantum stage was approximately 0.54 GB. The ResNet experiments required more memory because they operate directly on image tensors using an 11.2M-parameter CNN. The maximum recorded resident-set peak was 2.24 GB for RGB input and 2.71 GB for the 13-band multispectral model.

The wall times in Table~\ref{tab:runtime_main} also explain why the ablation studies were staged. A complete architecture grid spanning multiple circuit depths, entangling topologies, PCA dimensions, and qubit counts would multiply the cost of the 225-job quantum stage many times over. The representative 20-pair subset is therefore used for exploratory architecture and qubit-count studies, while the final PCA-32 configuration is evaluated on all 45 class pairs. This strategy keeps model selection computationally feasible while ensuring that the final performance is reported on the complete pairwise benchmark.

\subsection{\texorpdfstring{Scope and Future Directions}{Limitations}}

The present study focuses on understanding learned quantum representations under a controlled experimental protocol. The quantum models are evaluated in noiseless simulation, the PQC architecture is selected from a representative rather than exhaustive search, and the PCA-based workflow deliberately prioritizes compact spectral representations over direct image-level processing. Likewise, the external datasets are evaluated using the fixed EuroSAT-MS configuration to assess transfer rather than to maximize performance on each dataset individually. These design choices allow controlled comparisons between quantum and classical readouts while keeping the experimental study computationally tractable.

Future work can naturally extend these results in several directions. Larger quantum processors and improved simulation methods may enable direct processing of high-dimensional spectral-spatial data without aggressive classical dimensionality reduction. Broader architectural searches, hardware-aware training, multimodal Sentinel-1/Sentinel-2 representations, and more scalable kernel approximations may further improve both performance and practicality. The projected-kernel experiments presented here provide an initial step toward such scalable quantum representations and motivate further investigation of representation and readout design.

\FloatBarrier

\section{Conclusions}

This work presented a systematic study of PQCs for multispectral land-cover classification under a controlled experimental protocol. By combining pairwise classification, validation-only model selection, statistical testing, and multiple dataset evaluations, we showed that compact PQCs can learn meaningful nonlinear representations of multispectral data. The experiments further demonstrate that the learned representation and its downstream readout should be regarded as distinct components: the same trained circuit can support different classification strategies whose effectiveness depends on the geometry of the learned feature space and the target dataset. While the trained quantum kernel provides consistent improvements over the local VQC readout on EuroSAT-MS, no single readout is universally preferable, emphasizing that representation learning and readout design should be considered jointly in QML.

The present results should be viewed as a step toward understanding how quantum representations can be learned, interpreted, and ultimately deployed for remote sensing. As quantum processors continue to scale and more efficient kernel estimation and representation techniques become available, future quantum models may move beyond compact PCA representations toward direct processing of high-dimensional spectral and spectral--spatial information. More broadly, this work suggests that the future development of QML will depend not only on increasing quantum resources, but also on understanding how to construct representations and readouts that exploit those resources effectively.

\FloatBarrier

\appendices

\section{\texorpdfstring{Representative Pair Subset}{Representative Pair Subset}}

\label{app:representative_pairs}

The representative 20-pair subset used for architecture and qubit-count studies was selected deterministically from the 45 one-vs-one pairs using validation accuracy only. Each pair received a composite difficulty score equal to the average percentile rank of logistic regression, SVM-linear, SVM-RBF, and NN validation accuracy. The 45 pairs were divided into five difficulty quintiles, and four pairs were selected from each quintile. The selection was constrained so that each of the ten classes appeared in exactly four selected pairs. The selected pairs were:

\begin{center}

\tiny
\setlength{\tabcolsep}{1.5pt}
\begin{tabular}{@{}ll@{}}
\hline
\multicolumn{2}{c}{Quintile-balanced selected pairs} \\
\hline
AnnualCrop--PermanentCrop & Highway--Residential \\
Highway--Pasture & HerbaceousVegetation--Residential \\
HerbaceousVegetation--PermanentCrop & AnnualCrop--Residential \\
AnnualCrop--Pasture & Forest--Highway \\
Residential--River & Pasture--River \\
AnnualCrop--River & Industrial--PermanentCrop \\
Forest--River & Forest--HerbaceousVegetation \\
Industrial--Pasture & Forest--Industrial \\
Highway--SeaLake & Industrial--SeaLake \\
HerbaceousVegetation--SeaLake & PermanentCrop--SeaLake \\
\hline
\end{tabular}
\end{center}

\FloatBarrier

\section{\texorpdfstring{Protocol and Selection Audit}{Protocol and Selection Audit}}

\label{app:protocol_audit}

Table~\ref{tab:protocol_audit} summarizes the main controls used to prevent leakage and post-hoc test-set selection. The same global split is used throughout the full PCA-32/PCA-64 evaluation. PCA bases and normalization statistics are fitted separately for each binary pair using training data only. The architecture and PCA choices are made from validation results and cost considerations; test accuracy is reserved for the final comparisons in Section~\ref{sec:results}.

\begin{table}[!t]

\centering
\caption{Protocol controls used in the revision experiments.}
\label{tab:protocol_audit}
\footnotesize
\begin{tabular}{p{0.36\linewidth} p{0.55\linewidth}}
\hline
Item & Control \\
\hline
Split identity & One fixed global split with 1400/300/300 images per class for train/validation/test. \\
Pairwise preprocessing & PCA, mean, and standard deviation fitted only on the pairwise training partition. \\
Seed variation & Seeds change initialization and optimization order, not image membership. \\
PCA selection & PCA chosen using validation accuracy, explained variance, feature coverage, and runtime. \\
Circuit selection & \(L\), topology, and offset chosen using validation results on a deterministic representative subset. \\
Test use & Test accuracy used only after PCA and circuit choices are fixed. \\
OvO evaluation & All pairwise classifiers score the identical global test image identifiers before vote aggregation. \\
Gram storage & Gram matrices are not saved for the full final runs; metrics and predictions are saved. \\
\hline
\end{tabular}
\end{table}

\FloatBarrier

\section{\texorpdfstring{Class-Level Confidence Intervals}{Class-Level Confidence Intervals}}

\label{app:class_ci}

For the appendix tables, accuracy is first averaged over the nine one-vs-one pairs involving each class for a fixed training seed. The 95\% confidence-interval half-width is then computed over the five seeds as

\begin{equation}
\mathrm{CI}_{95}=t_{0.975,n-1}\frac{s}{\sqrt{n}},
\quad n=5,\quad t_{0.975,4}=2.776,
\end{equation}

where \(s\) is the sample standard deviation across seeds. Values are reported as mean\(\pm\)CI half-width in percentage points. The macro-average row reports the mean of the per-class means; its \(\pm\) value is the average per-class CI half-width and is not an independent confidence interval for the macro-average. Because the final split is fixed, deterministic models such as logistic regression, SVM-linear, and SVM-RBF have zero seed-level CI; their across-task variability is instead reflected by the class-pair means and paired statistical tests in the main text.

\begin{table}[H]

\centering
\caption{Per-class mean\(\pm\)95\% CI (\%) over five seeds for logistic regression, parameter-matched NN, and VQC.}
\label{tab:app_ci_logreg_nn_vqc}
\scriptsize
\setlength{\tabcolsep}{2.5pt}
\begin{tabular}{l c c c}
\hline
Class & LogReg & NN & VQC \\
\hline
AnnualCrop & 89.44\(\pm\)0.00 & 93.82\(\pm\)0.28 & 93.94\(\pm\)0.28 \\
Forest & 97.52\(\pm\)0.00 & 98.53\(\pm\)0.12 & 98.63\(\pm\)0.17 \\
HerbaceousVegetation & 94.28\(\pm\)0.00 & 95.59\(\pm\)0.36 & 95.29\(\pm\)0.35 \\
Highway & 86.04\(\pm\)0.00 & 89.77\(\pm\)0.39 & 89.74\(\pm\)0.40 \\
Industrial & 94.87\(\pm\)0.00 & 96.05\(\pm\)0.15 & 95.98\(\pm\)0.21 \\
Pasture & 94.44\(\pm\)0.00 & 95.92\(\pm\)0.33 & 96.33\(\pm\)0.18 \\
PermanentCrop & 91.22\(\pm\)0.00 & 93.67\(\pm\)0.35 & 94.18\(\pm\)0.38 \\
Residential & 91.02\(\pm\)0.00 & 94.13\(\pm\)0.27 & 94.17\(\pm\)0.35 \\
River & 93.41\(\pm\)0.00 & 96.23\(\pm\)0.38 & 96.07\(\pm\)0.35 \\
SeaLake & 99.76\(\pm\)0.00 & 99.73\(\pm\)0.06 & 99.80\(\pm\)0.08 \\
Macro-average & 93.20\(\pm\)0.00 & 95.35\(\pm\)0.27 & 95.41\(\pm\)0.27 \\
\hline
\end{tabular}
\end{table}

\begin{table}[H]

\centering
\caption{Per-class mean\(\pm\)95\% CI (\%) over five seeds for SVM-linear, SVM-QK (trained), and SVM-RBF.}
\label{tab:app_ci_svm}
\scriptsize
\setlength{\tabcolsep}{2.5pt}
\begin{tabular}{l c c c}
\hline
Class & SVM-linear & SVM-QK (trained) & SVM-RBF \\
\hline
AnnualCrop & 90.06\(\pm\)0.00 & 94.52\(\pm\)0.28 & 95.20\(\pm\)0.00 \\
Forest & 97.69\(\pm\)0.00 & 98.91\(\pm\)0.14 & 98.83\(\pm\)0.00 \\
HerbaceousVegetation & 94.30\(\pm\)0.00 & 95.67\(\pm\)0.44 & 96.22\(\pm\)0.00 \\
Highway & 86.15\(\pm\)0.00 & 90.18\(\pm\)0.26 & 91.87\(\pm\)0.00 \\
Industrial & 94.81\(\pm\)0.00 & 96.28\(\pm\)0.16 & 96.69\(\pm\)0.00 \\
Pasture & 94.54\(\pm\)0.00 & 96.79\(\pm\)0.37 & 97.04\(\pm\)0.00 \\
PermanentCrop & 91.63\(\pm\)0.00 & 94.69\(\pm\)0.28 & 95.30\(\pm\)0.00 \\
Residential & 90.96\(\pm\)0.00 & 94.76\(\pm\)0.26 & 95.91\(\pm\)0.00 \\
River & 93.31\(\pm\)0.00 & 96.58\(\pm\)0.18 & 97.80\(\pm\)0.00 \\
SeaLake & 99.78\(\pm\)0.00 & 99.85\(\pm\)0.02 & 99.19\(\pm\)0.00 \\
Macro-average & 93.32\(\pm\)0.00 & 95.82\(\pm\)0.24 & 96.40\(\pm\)0.00 \\
\hline
\end{tabular}
\end{table}

\begin{table}[H]

\centering
\caption{Per-class mean\(\pm\)95\% CI (\%) over five seeds comparing VQC with SVM-QK (trained) from the same circuit.}
\label{tab:app_ci_vqc_qk}
\scriptsize
\setlength{\tabcolsep}{2.5pt}
\begin{tabular}{l c c}
\hline
Class & VQC & SVM-QK (trained) \\
\hline
AnnualCrop & 93.94\(\pm\)0.28 & 94.52\(\pm\)0.28 \\
Forest & 98.63\(\pm\)0.17 & 98.91\(\pm\)0.14 \\
HerbaceousVegetation & 95.29\(\pm\)0.35 & 95.67\(\pm\)0.44 \\
Highway & 89.74\(\pm\)0.40 & 90.18\(\pm\)0.26 \\
Industrial & 95.98\(\pm\)0.21 & 96.28\(\pm\)0.16 \\
Pasture & 96.33\(\pm\)0.18 & 96.79\(\pm\)0.37 \\
PermanentCrop & 94.18\(\pm\)0.38 & 94.69\(\pm\)0.28 \\
Residential & 94.17\(\pm\)0.35 & 94.76\(\pm\)0.26 \\
River & 96.07\(\pm\)0.35 & 96.58\(\pm\)0.18 \\
SeaLake & 99.80\(\pm\)0.08 & 99.85\(\pm\)0.02 \\
Macro-average & 95.41\(\pm\)0.27 & 95.82\(\pm\)0.24 \\
\hline
\end{tabular}
\end{table}

\FloatBarrier

\FloatBarrier

\section{\texorpdfstring{Full Configuration-Search Grid}{Full Configuration-Search Grid}}

\label{app:config_grid}

Table~\ref{tab:config_search_full} gives the full representative-subset configuration grid summarized graphically in Fig.~\ref{fig:config_search}. PCA-8 is not included in this grid because it was screened out by the 45-pair PCA sensitivity analysis in Table~\ref{tab:pca_sensitivity}.

\begin{table}[!h]

\centering
\caption{Full PCA--circuit configuration grid on the 20 representative EuroSAT-MS pairs. Values are validation accuracy (\%) averaged over pair means and five training seeds.}
\label{tab:config_search_full}
\scriptsize
\resizebox{\linewidth}{!}{%
\begin{tabular}{c c c l c c c c}
\hline
PCA & Offset & \(L\) & Topology & Params & CNOTs & VQC & SVM-QK \\
\hline
16 & 7  & 2  & alternating &  61 &  7 & 94.90 & 94.65 \\
16 & 7  & 2  & ring        &  61 &  8 & 94.99 & 94.86 \\
16 & 7  & 4  & alternating & 109 & 14 & 95.68 & 95.43 \\
16 & 7  & 4  & ring        & 109 & 16 & 95.70 & 95.56 \\
16 & 7  & 6  & alternating & 157 & 21 & 95.85 & 95.61 \\
16 & 7  & 6  & ring        & 157 & 24 & 95.79 & 95.56 \\
16 & 7  & 8  & alternating & 205 & 28 & 95.85 & 95.71 \\
16 & 7  & 8  & ring        & 205 & 32 & 95.87 & 95.76 \\
\hline
32 & 8  & 2  & alternating &  61 &  7 & 95.12 & 95.09 \\
32 & 8  & 2  & ring        &  61 &  8 & 95.11 & 94.98 \\
32 & 8  & 4  & alternating & 109 & 14 & 96.08 & 95.86 \\
32 & 8  & 4  & ring        & 109 & 16 & 96.06 & 95.85 \\
32 & 8  & 6  & alternating & 157 & 21 & 96.16 & 95.91 \\
32 & 8  & 6  & ring        & 157 & 24 & 96.16 & 95.92 \\
32 & 8  & 8  & alternating & 205 & 28 & 96.24 & 96.09 \\
32 & 8  & 8  & ring        & 205 & 32 & 96.34 & 96.23 \\
32 & 8  & 10 & alternating & 253 & 35 & 96.38 & 96.28 \\
32 & 8  & 10 & ring        & 253 & 40 & 96.41 & 96.26 \\
32 & 8  & 12 & alternating & 301 & 42 & 96.42 & 96.19 \\
32 & 8  & 12 & ring        & 301 & 48 & 96.43 & 96.20 \\
\hline
64 & 32 & 8  & alternating & 205 & 28 & 96.28 & 96.07 \\
64 & 32 & 10 & alternating & 253 & 35 & 96.50 & 96.46 \\
64 & 32 & 12 & alternating & 301 & 42 & 96.49 & 96.33 \\
64 & 32 & 14 & alternating & 349 & 49 & 96.56 & 96.45 \\
\hline
\end{tabular}
}
\end{table}

\FloatBarrier

\section{\texorpdfstring{PCA-64 Sensitivity}{PCA-64 Sensitivity}}

\label{app:pca64}

PCA-64 was evaluated under the final \(L=10\) alternating circuit using offset 32 so that all 64 components entered the reuploading schedule. Table~\ref{tab:pca32_pca64_full} gives the full model-by-model comparison. The linear models benefit from PCA-64 because the additional components preserve more variance that a linear boundary can use directly. For VQC, SVM-QK, NN, and SVM-RBF, the additional components do not translate into a statistically meaningful validation gain over PCA-32.

\begin{table}[!h]

\centering
\caption{Full PCA-32 versus PCA-64 comparison over all 45 class pairs. Values are means over class-pair means. The NN values in this table correspond to the compact NN used in the PCA sensitivity runs.}
\label{tab:pca32_pca64_full}
\footnotesize
\begin{tabular}{l c c c c}
\hline
Model & \multicolumn{2}{c}{PCA-32} & \multicolumn{2}{c}{PCA-64} \\
 & Val. & Test & Val. & Test \\
\hline
Logistic regression & 92.77 & 93.20 & \textbf{93.62} & \textbf{94.29} \\
SVM-linear          & 92.84 & 93.32 & \textbf{93.73} & \textbf{94.35} \\
Compact NN          & \textbf{95.98} & \textbf{95.35} & 95.88 & 95.26 \\
VQC                 & 96.18 & \textbf{95.41} & \textbf{96.26} & 95.39 \\
SVM-QK (trained)    & 96.01 & 95.82 & \textbf{96.07} & \textbf{95.95} \\
SVM-RBF             & \textbf{96.10} & \textbf{96.40} & 96.08 & 96.37 \\
\hline
\end{tabular}
\end{table}

\FloatBarrier

\FloatBarrier

\section{\texorpdfstring{Additional Qubit-Sweep Values}{Additional Qubit-Sweep Values}}

\label{app:qubit_values}

Table~\ref{tab:qubit_values} gives the numerical values behind Figs.~\ref{fig:qubit_vqc_classical} and~\ref{fig:qubit_kernel_readouts}. The projected kernels use one-local \(X,Y,Z\) features and edge two-local Pauli features, followed by either a linear kernel or an RBF kernel with the standard scale gamma rule.

\begin{table}[!h]

\centering
\caption{Qubit-count sweep on 20 representative pairs. Values are test accuracy (\%).}
\label{tab:qubit_values}
\scriptsize
\setlength{\tabcolsep}{1.8pt}
\begin{tabular}{c c c c c c}
\hline
\(q\) & VQC & Param.-matched NN & SVM-QK & Proj. QK-lin. & Proj. QK-RBF \\
\hline
3  & 95.59 & 95.39 & 96.02 & 95.86 & 96.13 \\
4  & 95.78 & 95.82 & 96.34 & 96.15 & 96.42 \\
5  & 95.79 & 95.89 & 96.53 & 96.13 & 96.44 \\
7  & 96.16 & 96.21 & \textbf{96.77} & 96.35 & 96.58 \\
9  & 96.30 & 96.44 & 96.11 & \textbf{96.51} & 96.59 \\
11 & \textbf{96.32} & \textbf{96.57} & 95.10 & 96.49 & \textbf{96.64} \\
\hline
\end{tabular}
\end{table}

\FloatBarrier

\FloatBarrier

\section{\texorpdfstring{Runtime and Memory}{Runtime and Memory}}

\label{app:runtime}

All timings in Table~\ref{tab:runtime_main} were measured on the same CPU-only system. The PCA-32 quantum run completed all 225 pair-seed jobs. The mean VQC training time was 268.6 s per pair-seed, and the mean complete quantum job time, including cached trained-kernel construction and SVM fitting, was 273.5 s. Kernel construction and SVM fitting added only 1.6 s per pair-seed in simulation because trained state vectors were cached and Gram matrices were computed by dense linear algebra. This timing should not be extrapolated directly to hardware, where each kernel entry would require circuit execution and finite shots.

The parameter-matched NN run used the same pairwise data splits and completed 225 pair-seed jobs in 1.15 h with eight workers. Its mean training time was 140.7 s per pair-seed. The deterministic SVM and logistic-regression baselines were much cheaper: in the recorded PCA-32 stage, SVM-linear required about 0.17 s per pair-seed and SVM-RBF about 0.24 s per pair-seed. The direct image ResNet runs were slower because they train 11M-parameter CNNs on 14000 images per seed; the recorded full wall times were 7.62 h for RGB and 8.80 h for 13-band multispectral ResNet-18.

The final PCA-32/PCA-64 full driver was run sequentially by stage but parallelized within the pair-seed stages. Its total wall time was 9.50 h. The stage-level wall times were 3.22 h for PCA-32 quantum, 1.71 h for PCA-32 classical, 0.62 min for PCA-32 OvO aggregation, 3.11 h for PCA-64 quantum, 1.43 h for PCA-64 classical, and 0.25 min for PCA-64 OvO aggregation. The OvO stage is fast because it reuses saved pairwise predictions and only performs vote aggregation. The quantum and NN stages dominate runtime because they require iterative training.

Peak memory was not a limiting factor in these CPU runs. The largest recorded PCA-32 quantum job resident-set peak was about 0.54 GB. The ResNet baselines used more memory, with maximum recorded resident-set peaks of 2.24 GB for RGB and 2.71 GB for 13-band multispectral input. Disk usage was kept moderate by not saving full Gram matrices in the final PCA-32/PCA-64 runs; the saved artifacts consist mainly of manifests, metrics, checkpoints, and predictions.

\FloatBarrier

\bibliographystyle{IEEEtran}
\bibliography{references}

@article{helber2019eurosat,
  title={EuroSAT: A novel dataset and deep learning benchmark for land use and land cover classification},
  author={Helber, Patrick and Bischke, Benjamin and Dengel, Andreas and Borth, Damian},
  journal={IEEE Journal of Selected Topics in Applied Earth Observations and Remote Sensing},
  volume={12},
  number={7},
  pages={2217--2226},
  year={2019},
  publisher={IEEE}
}

@article{schuld2021advantages,
  title={The advantages and bottlenecks of quantum machine learning},
  author={Schuld, Maria and Sweke, Ryan and Meyer, Johannes Jakob and Gross, David and Petruccione, Francesco},
  journal={arXiv preprint arXiv:2101.10657},
  year={2021}
}

@article{abbas2021universal,
  title={Universal expressiveness of variational quantum classifiers and quantum kernels for support vector machines},
  author={Abbas, Alaa and Diamanti, Eleni and Wehner, Stephanie},
  journal={Nature Communications},
  volume={14},
  number={1},
  pages={530},
  year={2023},
  doi = {10.1038/s41467-023-36144-5}
}

@article{Fan2024Land,
  author={Fan, Fan and Shi, Yilei and Zhu, Xiao Xiang},
  title={Land Cover Classification From Sentinel-2 Images With Quantum-Classical Convolutional Neural Networks},
  journal={IEEE Journal of Selected Topics in Applied Earth Observations and Remote Sensing},
  year={2024},
  volume={17},
  pages={12477--12489},
  doi={10.1109/JSTARS.2024.3411670}
}

@article{ma2019dl_remote_sensing_review,
  title   = {Deep learning in remote sensing applications: A meta-analysis and review},
  author  = {Ma, Lei and Liu, Yu and Zhang, Xueliang and Ye, Yuanxin and Yin, Gaofei and Johnson, Brian Alan},
  journal = {ISPRS Journal of Photogrammetry and Remote Sensing},
  volume  = {152},
  pages   = {166--177},
  year    = {2019},
  doi     = {10.1016/j.isprsjprs.2019.04.015}
}

@article{schuld2019feature_hilbert_spaces,
  title     = {Quantum Machine Learning in Feature Hilbert Spaces},
  author    = {Schuld, Maria and Killoran, Nathan},
  journal   = {Physical Review Letters},
  volume    = {122},
  number    = {4},
  pages     = {040504},
  year      = {2019},
  doi       = {10.1103/PhysRevLett.122.040504},
  publisher = {American Physical Society}
}

@article{havlivcek2019quantum_enhanced_feature_spaces,
  title     = {Supervised learning with quantum-enhanced feature spaces},
  author    = {Havl{\'i}{\v{c}}ek, Vojt{\v{e}}ch and C{\'o}rcoles, Antonio D. and Temme, Kristan and Harrow, Aram W. and Kandala, Abhinav and Chow, Jerry M. and Gambetta, Jay M.},
  journal   = {Nature},
  volume    = {567},
  number    = {7747},
  pages     = {209--212},
  year      = {2019},
  doi       = {10.1038/s41586-019-0980-2},
  publisher = {Springer Nature}
}

@misc{schuld2021qml_kernel_methods,
  title         = {Supervised quantum machine learning models are kernel methods},
  author        = {Schuld, Maria},
  year          = {2021},
  eprint        = {2101.11020},
  archivePrefix = {arXiv},
  primaryClass  = {quant-ph}
}

@article{schuld2021data_encoding_expressive_power,
  title     = {Effect of data encoding on the expressive power of variational quantum-machine-learning models},
  author    = {Schuld, Maria and Sweke, Ryan and Meyer, Johannes Jakob},
  journal   = {Physical Review A},
  volume    = {103},
  number    = {3},
  pages     = {032430},
  year      = {2021},
  doi       = {10.1103/PhysRevA.103.032430},
  publisher = {American Physical Society}
}

@article{cerezo2021vqa_review,
  title     = {Variational quantum algorithms},
  author    = {Cerezo, Marco and Arrasmith, Andrew and Babbush, Ryan and Benjamin, Simon C. and Endo, Suguru and Fujii, Keisuke and McClean, Jarrod R. and Mitarai, Kosuke and Yuan, Xiao and Cincio, Lukasz and Coles, Patrick J.},
  journal   = {Nature Reviews Physics},
  volume    = {3},
  number    = {9},
  pages     = {625--644},
  year      = {2021},
  doi       = {10.1038/s42254-021-00348-9},
}

@inproceedings{Zaidenberg2021Advantages,
  title={Advantages and Bottlenecks of Quantum Machine Learning for Remote Sensing},
  author={Zaidenberg, Daniela A. and Sebastianelli, A. and Spiller, D. and Ullo, Silvia Liberata},
  booktitle={2021 IEEE International Geoscience and Remote Sensing Symposium (IGARSS)},
  pages={5680--5683},
  year={2021},
  doi={10.1109/IGARSS47720.2021.9553133}
}

@article{Miroszewski2023Cloud,
  title={Detecting Clouds in Multispectral Satellite Images Using Quantum-Kernel Support Vector Machines},
  author={Miroszewski, Artur and Mielczarek, Jakub and Czelusta, Grzegorz and Szczepanek, Filip and Grabowski, Bartosz and Le Saux, Bertrand and Nalepa, Jakub},
  journal={IEEE Journal of Selected Topics in Applied Earth Observations and Remote Sensing},
  volume={16},
  pages={7601--7613},
  year={2023},
  doi={10.1109/JSTARS.2023.3304122},
  publisher={IEEE}
}

@inproceedings{Delilbasic2021Quantum,
  title={Quantum Support Vector Machine Algorithms for Remote Sensing Data Classification},
  author={Delilbasic, Amer and Cavallaro, Gabriele and Willsch, Marcel and Melgani, Farid and Riedel, Marc and Michielsen, Kristel},
  booktitle={2021 IEEE International Geoscience and Remote Sensing Symposium (IGARSS)},
  pages={2608--2611},
  year={2021},
  doi={10.1109/IGARSS47720.2021.9554802}
}

@article{Delilbasic2023A,
  title={A Single-Step Multiclass SVM Based on Quantum Annealing for Remote Sensing Data Classification},
  author={Delilbasic, Amer and Le Saux, Bertrand and Riedel, Marc and Michielsen, Kristel and Cavallaro, Gabriele},
  journal={IEEE Journal of Selected Topics in Applied Earth Observations and Remote Sensing},
  volume={17},
  pages={1434--1445},
  year={2024},
  doi={10.1109/JSTARS.2023.3336926}
}

@article{Sebastianelli2021On,
  title={On Circuit-Based Hybrid Quantum Neural Networks for Remote Sensing Imagery Classification},
  author={Sebastianelli, Alessandro and Zaidenberg, Daniela A. and Spiller, Dario and Le Saux, Bertrand and Ullo, Silvia Liberata},
  journal={IEEE Journal of Selected Topics in Applied Earth Observations and Remote Sensing},
  volume={15},
  pages={565--580},
  year={2022},
  doi={10.1109/JSTARS.2021.3134785}
}

@article{Fan2023Hybrid,
  title={Hybrid Quantum-Classical Convolutional Neural Network Model for Image Classification},
  author={Fan, Fan and Shi, Yilei and Guggemos, Tobias and Zhu, Xiao Xiang},
  journal={IEEE Transactions on Neural Networks and Learning Systems},
  volume={35},
  number={12},
  pages={18145--18159},
  year={2024},
  doi={10.1109/TNNLS.2023.3312170}
}

@article{Liliopoulos2025Hybrid,
  title={Hybrid classical-quantum multilayer neural networks for monitoring agricultural activities using remote sensing data},
  author={Liliopoulos, Ioannis and Varsamis, Georgios D. and Milchanowski, Kristin and Martin-Cuevas, Rafael and Safouri, Konstantina and Dimitrakis, Panagiotis and Karafyllidis, Ioannis G.},
  journal={Quantum Machine Intelligence},
  volume={7},
  year={2025},
  doi={10.1007/s42484-024-00230-8}
}

@article{Sebastianelli2024Quanv4EO,
  title={Quanv4EO: Empowering Earth Observation by Means of Quanvolutional Neural Networks},
  author={Sebastianelli, Alessandro and Mauro, Francesco and Ciabatti, Giulia and Spiller, Dario and Le Saux, Bertrand and Gamba, Paolo and Ullo, Silvia Liberata},
  journal={IEEE Transactions on Geoscience and Remote Sensing},
  volume={63},
  pages={1--15},
  year={2025},
  doi={10.1109/TGRS.2025.3556335}
}

@article{Sebastianelli2023On,
  title={On Quantum Hyperparameters Selection in Hybrid Classifiers for Earth Observation Data},
  author={Sebastianelli, Alessandro and Di Rosso, Maria P. and Ullo, Silvia Liberata and Gamba, Paolo},
  journal={IEEE Geoscience and Remote Sensing Letters},
  volume={20},
  pages={1--5},
  year={2023},
  doi={10.1109/LGRS.2023.3308105}
}

@article{Otgonbaatar2023Exploiting,
  title={Exploiting the Quantum Advantage for Satellite Image Processing: Review and Assessment},
  author={Otgonbaatar, Soronzonbold and Kranzlm{\"u}ller, Dieter},
  journal={IEEE Transactions on Quantum Engineering},
  volume={5},
  pages={1--9},
  year={2024},
  doi={10.1109/TQE.2023.3338970}
}

@article{Liu2023Quantum,
  title={Quantum Machine Learning on Remote Sensing Data Classification},
  author={Liu, Yi and Wang, Wendy and Wang, Haibo and Alidaee, B.},
  journal={Journal of Engineering Research and Sciences},
  year={2023},
  doi={10.55708/js0212004}
}

@article{Shaik2022Quantum,
  title={Quantum Based Pseudo-Labelling for Hyperspectral Imagery: A Simple and Efficient Semi-Supervised Learning Method for Machine Learning Classifiers},
  author={Shaik, R. and Unni, Aiswarya and Zeng, Weiping},
  journal={Remote Sensing},
  volume={14},
  number={22},
  pages={5774},
  year={2022},
  doi={10.3390/rs14225774}
}

@inproceedings{Zollner2022Quantum,
  title={Quantum classifiers for remote sensing},
  author={Zollner, Johann Maximilian},
  booktitle={Proceedings of the 30th International Conference on Advances in Geographic Information Systems},
  year={2022},
  doi={10.1145/3557915.3565537}
}

@article{Benedetti2019PQC,
  title={Parameterized quantum circuits as machine learning models},
  author={Benedetti, Marcello and Lloyd, Erika and Sack, Stefan and Fiorentini, Mattia},
  journal={Quantum Science and Technology},
  volume={4},
  number={4},
  pages={043001},
  year={2019},
  doi={10.1088/2058-9565/ab4eb5}
}

@article{PerezSalinas2020DataReuploading,
  title     = {Data re-uploading for a universal quantum classifier},
  author    = {P{\'e}rez-Salinas, Adri{\'a}n and Cervera-Lierta, Alba and Gil-Fuster, Elies and Latorre, Jos{\'e} I.},
  journal   = {Quantum},
  volume    = {4},
  pages     = {226},
  year      = {2020},
  doi       = {10.22331/q-2020-02-06-226}
}

@article{Sim2019Expressibility,
  title     = {Expressibility and Entangling Capability of Parameterized Quantum Circuits for Hybrid Quantum-Classical Algorithms},
  author    = {Sim, Sukin and Johnson, Peter D. and Aspuru-Guzik, Al{\'a}n},
  journal   = {Advanced Quantum Technologies},
  volume    = {2},
  number    = {12},
  pages     = {1900070},
  year      = {2019},
  doi       = {10.1002/qute.201900070}
}

@article{mcclean2018barren,
  title     = {Barren plateaus in quantum neural network training landscapes},
  author    = {McClean, Jarrod R. and Boixo, Sergio and Smelyanskiy, Vadim N. and Babbush, Ryan and Neven, Hartmut},
  journal   = {Nature Communications},
  volume    = {9},
  number    = {1},
  pages     = {4812},
  year      = {2018},
  doi       = {10.1038/s41467-018-07090-4},
  publisher = {Nature Publishing Group}
}

@article{maragkopoulos2026quantum,
  title={Quantum-Inspired Unitary Pooling for Multispectral Satellite Image Classification},
  author={Maragkopoulos, Georgios and Mandilara, Aikaterini and Komini, Ralntion and Syvridis, Dimitris},
  journal={arXiv preprint arXiv:2603.15522},
  year={2026}
}

@article{katerina,
  author  = {Mandilara, Aikaterini and Papadopoulos, Aristeides and Syvridis, D.},
  title   = {Learning quantum kernels with continuous-variable optical circuits},
  journal = {Machine Learning: Science and Technology},
  year    = {2026},
  doi     = {10.1088/2632-2153/ae7baf},
  publisher = {IOP Publishing}
}

@article{Hubregtsen,
  title = {Training quantum embedding kernels on near-term quantum computers},
  author = {Hubregtsen, Thomas and Wierichs, David and Gil-Fuster, Elies and Derks, Peter-Jan H. S. and Faehrmann, Paul K. and Meyer, Johannes Jakob},
  journal = {Phys. Rev. A},
  volume = {106},
  issue = {4},
  pages = {042431},
  numpages = {18},
  year = {2022},
  month = {Oct},
  publisher = {American Physical Society},
  doi = {10.1103/PhysRevA.106.042431},
  url = {https://link.aps.org/doi/10.1103/PhysRevA.106.042431}
}

@inproceedings{sumbul2019bigearthnet,
  title={BigEarthNet: A Large-Scale Benchmark Archive for Remote Sensing Image Understanding},
  author={Sumbul, Gencer and Charfuelan, Marcela and Demir, Begum and Markl, Volker},
  booktitle={2019 IEEE International Geoscience and Remote Sensing Symposium (IGARSS)},
  pages={5901--5904},
  year={2019},
  doi={10.1109/IGARSS.2019.8900532}
}

@article{khan2024transformer_lulc,
  title={Transformer-based land use and land cover classification with explainability using satellite imagery},
  author={Khan, Mehak and Hanan, Abdul and Kenzhebay, Meruyert and Gazzea, Michele and Arghandeh, Reza},
  journal={Scientific Reports},
  volume={14},
  pages={16744},
  year={2024},
  doi={10.1038/s41598-024-67186-4}
}

@inproceedings{helber2018introducing,
  title={Introducing EuroSAT: A Novel Dataset and Deep Learning Benchmark for Land Use and Land Cover Classification},
  author={Helber, Patrick and Bischke, Benjamin and Dengel, Andreas and Borth, Damian},
  booktitle={IGARSS 2018-2018 IEEE International Geoscience and Remote Sensing Symposium},
  pages={204--207},
  year={2018},
  organization={IEEE}
}

@inproceedings{cong2022satmae,
  title={SatMAE: Pre-training Transformers for Temporal and Multi-Spectral Satellite Imagery},
  author={Cong, Yezhen and Khanna, Samar and Meng, Chenlin and Liu, Patrick and Rozi, Erik and He, Yutong and Burke, Marshall and Lobell, David B. and Ermon, Stefano},
  booktitle={Advances in Neural Information Processing Systems},
  volume={35},
  pages={197--211},
  year={2022},
  doi = {10.52202/068431-0015}
}

@article{szwarcman2025prithvi,
  title={Prithvi-eo-2.0: A versatile multi-temporal foundation model for earth observation applications},
  author={Szwarcman, Daniela and Roy, Sujit and Fraccaro, Paolo and G{\'\i}slason, Orsteinn El{\'\i} and Blumenstiel, Benedikt and Ghosal, Rinki and De Oliveira, Pedro Henrique and de Sousa Almeida, Joao Lucas and Sedona, Rocco and Kang, Yanghui and others},
  journal={IEEE Transactions on Geoscience and Remote Sensing},
  year={2025},
  publisher={IEEE},
  doi = {10.1109/TGRS.2025.3642610}
}

@article{wang2021noise_induced,
  author  = {Wang, Samson and Fontana, Enrico and Cerezo, M. and
             Sharma, Kunal and Sone, Akira and Cincio, {\L}ukasz and
             Coles, Patrick J.},
  title   = {Noise-Induced Barren Plateaus in Variational Quantum Algorithms},
  journal = {Nature Communications},
  volume  = {12},
  number  = {1},
  pages   = {6961},
  year    = {2021},
  doi     = {10.1038/s41467-021-27045-6},
  url     = {https://doi.org/10.1038/s41467-021-27045-6}
}

@article{thanasilp2024concentration,
  author  = {Thanasilp, Supanut and Wang, Samson and Cerezo, M. and
             Holmes, Zo{\"e}},
  title   = {Exponential Concentration in Quantum Kernel Methods},
  journal = {Nature Communications},
  volume  = {15},
  number  = {1},
  pages   = {5200},
  year    = {2024},
  doi     = {10.1038/s41467-024-49287-w},
  url     = {https://doi.org/10.1038/s41467-024-49287-w}
}

@article{peters2021noisy_kernel,
  author  = {Peters, Evan and Caldeira, Jo{\~a}o and Ho, Alan and
             Leichenauer, Stefan and Mohseni, Masoud and Neven, Hartmut and
             Spentzouris, Panagiotis and Strain, Doug and Perdue, Gabriel N.},
  title   = {Machine Learning of High Dimensional Data on a Noisy Quantum Processor},
  journal = {npj Quantum Information},
  volume  = {7},
  number  = {1},
  pages   = {161},
  year    = {2021},
  doi     = {10.1038/s41534-021-00498-9},
  url     = {https://doi.org/10.1038/s41534-021-00498-9}
}

@misc{ibm2025heron,
  author       = {Mandelbaum, Ryan},
  title        = {Scaling for Quantum Advantage and Beyond},
  howpublished = {IBM Quantum Computing Blog},
  month        = nov,
  year         = {2025},
  url          = {https://www.ibm.com/quantum/blog/qdc-2025},
  note         = {Accessed: 2026-07-02}
}

@article{coelho2025quantumefficient,
  title   = {Quantum-Efficient Kernel Target Alignment},
  author  = {Coelho, Rodrigo and Kruse, Georg and Rosskopf, Andreas},
  journal = {arXiv preprint arXiv:2502.08225},
  year    = {2025},
  doi     = {10.48550/arXiv.2502.08225}
}

@article{haug2023randomized,
  title   = {Quantum Machine Learning of Large Datasets Using Randomized Measurements},
  author  = {Haug, Tobias and Self, Chris N. and Kim, M. S.},
  journal = {Machine Learning: Science and Technology},
  volume  = {4},
  number  = {1},
  pages   = {015005},
  year    = {2023},
  doi     = {10.1088/2632-2153/acb0b4}
}

@article{huang2021power,
  title   = {Power of Data in Quantum Machine Learning},
  author  = {Huang, Hsin-Yuan and Broughton, Michael and Mohseni, Masoud
             and Babbush, Ryan and Boixo, Sergio and Neven, Hartmut
             and McClean, Jarrod R.},
  journal = {Nature Communications},
  volume  = {12},
  pages   = {2631},
  year    = {2021},
  doi     = {10.1038/s41467-021-22539-9}
}

@article{jerbi2024shadows,
  title   = {Shadows of Quantum Machine Learning},
  author  = {Jerbi, Sofiene and Gyurik, Casper and Marshall, Simon C.
             and Molteni, Riccardo and Dunjko, Vedran},
  journal = {Nature Communications},
  volume  = {15},
  pages   = {5676},
  year    = {2024},
  doi     = {10.1038/s41467-024-49877-8}
}

@article{schuld2019gradients,
  title   = {Evaluating Analytic Gradients on Quantum Hardware},
  author  = {Schuld, Maria and Bergholm, Ville and Gogolin, Christian
             and Izaac, Josh and Killoran, Nathan},
  journal = {Physical Review A},
  volume  = {99},
  number  = {3},
  pages   = {032331},
  year    = {2019},
  doi     = {10.1103/PhysRevA.99.032331}
}

@article{zhu2020so2sat,
  title   = {{So2Sat LCZ42}: A Benchmark Data Set for the Classification of Global Local Climate Zones [Software and Data Sets]},
  author  = {Zhu, Xiao Xiang and Hu, Jingliang and Qiu, Chunping and Shi, Yilei
             and Kang, Jian and Mou, Lichao and Bagheri, Hossein
             and Haberle, Matthias and Hua, Yuansheng and Huang, Rong
             and Hughes, Lloyd and Li, Hao and Sun, Yao and Zhang, Guichen
             and Han, Shiyao and Schmitt, Michael and Wang, Yuanyuan},
  journal = {IEEE Geoscience and Remote Sensing Magazine},
  volume  = {8},
  number  = {3},
  pages   = {76--89},
  year    = {2020},
  doi     = {10.1109/MGRS.2020.2964708}
}

@inproceedings{basu2015deepsat,
  title     = {{DeepSat}: A Learning Framework for Satellite Imagery},
  author    = {Basu, Saikat and Ganguly, Sangram and Mukhopadhyay, Supratik
               and DiBiano, Robert and Karki, Manohar and Nemani, Ramakrishna},
  booktitle = {Proceedings of the 23rd SIGSPATIAL International Conference on Advances in Geographic Information Systems},
  series    = {SIGSPATIAL '15},
  articleno = {37},
  numpages  = {10},
  year      = {2015},
  publisher = {Association for Computing Machinery},
  doi       = {10.1145/2820783.2820816}
}
\end{document}